\documentclass[prl,aps,twocolumn,floatfix,superscriptaddress,preprintnumbers]{revtex4-1}
\usepackage{lmodern}
\usepackage{hyperref}
\hypersetup{colorlinks=true,linkcolor=purple,anchorcolor=blue,citecolor=blue, filecolor=blue,urlcolor=blue,bookmarksnumbered=true,
pdfview=FitB
}
\usepackage{mathrsfs}
\usepackage{rsfso}
\usepackage{color}
\usepackage{xcolor}
\usepackage{ulem}
\usepackage{csquotes}
\colorlet{purple1}{blue!70!red}
\colorlet{darkred}{red!50!black}

\usepackage{graphicx}
\usepackage[bf,SL,BF]{subfigure}
\usepackage{psfrag}
\usepackage{xcolor}
\usepackage{amssymb}
\usepackage{amsmath}
\usepackage{epstopdf}
\usepackage{natbib}
\newcommand{\be}{\begin{eqnarray}}
\newcommand{\ee}{\end{eqnarray}}

\begin{document}

\title{Towards a first principles light-front Hamiltonian for the nucleon}

\author{Siqi~Xu}
\email{xsq234@impcas.ac.cn} 
\affiliation{Institute of Modern Physics, Chinese Academy of Sciences, Lanzhou 730000, China}
\affiliation{School of Nuclear Science and Technology, University of Chinese Academy of Sciences, Beijing 100049, China}
\affiliation{Department of Physics and Astronomy, Iowa State University, Ames, Iowa 50011, USA}

\author{Yiping~Liu}
\email{liuyiping@impcas.ac.cn} 
\affiliation{Institute of Modern Physics, Chinese Academy of Sciences, Lanzhou 730000, China}
\affiliation{School of Nuclear Science and Technology, University of Chinese Academy of Sciences, Beijing 100049, China}

\author{Chandan~Mondal}
\email{mondal@impcas.ac.cn} 
\affiliation{Institute of Modern Physics, Chinese Academy of Sciences, Lanzhou 730000, China}
\affiliation{School of Nuclear Science and Technology, University of Chinese Academy of Sciences, Beijing 100049, China}

\author{Jiangshan~Lan}
\email{jiangshanlan@impcas.ac.cn} 
\affiliation{Institute of Modern Physics, Chinese Academy of Sciences, Lanzhou 730000, China}
\affiliation{School of Nuclear Science and Technology, University of Chinese Academy of Sciences, Beijing 100049, China}

\author{Xingbo~Zhao}
\email{xbzhao@impcas.ac.cn} 
\affiliation{Institute of Modern Physics, Chinese Academy of Sciences, Lanzhou 730000, China}
\affiliation{School of Nuclear Science and Technology, University of Chinese Academy of Sciences, Beijing 100049, China}

\author{Yang~Li}
\email{leeyoung1987@ustc.edu.cn} 
\affiliation{University of Science and Technology of China, Hefei, Anhui 230026, China}

\author{James~P.~Vary}
\email{jvary@iastate.edu} 
\affiliation{Department of Physics and Astronomy, Iowa State University, Ames, Iowa 50011, USA}

\collaboration{BLFQ Collaboration}

\date{\today}

\begin{abstract}

We solve the nucleon's wave functions from the eigenstates of the light-front quantum chromodynamics Hamiltonian for the first time, using a fully relativistic and nonperturbative approach based on light-front quantization, without an explicit confining potential. These eigenstates are determined for the three-quark, three-quark-gluon, and three-quark-quark-antiquark Fock representations, making them suitable for low-resolution probes. From this, we calculate the nucleon's quark and gluon matter densities, helicity, and transversity distributions, which show qualitative consistency with experimental extractions. We also compute the contributions of quark and gluon helicity to the proton spin and the tensor charges. The obtained light-front wave functions represent a significant advancement towards a unified description of various hadron distribution functions in both longitudinal and transverse momentum space.
  
\end{abstract}

\maketitle

{\it Introduction.}---
The visible world is built on nucleons (protons and neutrons), 
which form the atomic nuclei at the center of atoms that make up ordinary matter.
Understanding the formation of matter thus depends on explaining the dynamics and properties of the nucleons. Quantum chromodynamics (QCD) is the accepted theory for strong interactions~\cite{Callan:1977gz}, viewing nucleons as systems of confined quarks and gluons. However, predicting the extensive experimentally measured global static properties of nucleons—such as mass, spin, and size—remains challenging due to our limited understanding of nonperturbative QCD responsible for color confinement.

Successful theoretical frameworks for exploring some aspects of the partonic structure of hadrons are the discretized space-time Euclidean lattice~\cite{Hagler:2009ni,Joo:2019byq,MILC:2009mpl,BMW:2008jgk} and the Dyson-Schwinger equations of QCD~\cite{Maris:2003vk,Roberts:1994dr,Bashir:2012fs}. Significant progress is also being made within the Hamiltonian formulation of QCD quantized on the light front (LF)~\cite{Brodsky:1997de}. Additionally, light-front holography offers complementary insights into nonperturbative QCD~\cite{Brodsky:2014yha}. Basis light-front quantization (BLFQ), based on the Hamiltonian formalism, provides a nonperturbative framework for solving relativistic many-body bound state problems in quantum field theories such as QED and QCD~\cite{Vary:2009gt,Zhao:2014xaa,Nair:2022evk,Wiecki:2014ola,Li:2015zda,Jia:2018ary,Lan:2019vui,Mondal:2019jdg,Xu:2021wwj,Kuang:2022vdy,Lan:2021wok,Xu:2023nqv}. 
Until now, QCD applications of BLFQ to the mesons and baryons, such as the nucleons, have included a phenomenological confining interaction.

In this letter, we solve for the mass eigenstates of the LFQCD Hamiltonian using the BLFQ framework~\cite{Vary:2009gt} without the addition of a phenomenological confining interaction. With quarks ($q$) and gluons ($g$) as the explicit degrees of freedom, the Hamiltonian includes fundamental QCD interactions among them relevant to the constituent $|qqq\rangle$, $|qqqg\rangle$, and $|qqqq\bar{q}\rangle$ Fock sectors of the nucleon~\cite{Brodsky:1997de}. We compute the nucleon's electromagnetic form factors (FFs), parton distribution functions (PDFs), axial and tensor charges from the wave functions obtained as eigenvectors of the Hamiltonian. The Fourier transform of the FFs provides information about the spatial distributions of the nucleon’s constituents, such as charge and magnetization distributions. The PDFs encode the nonperturbative structure of the nucleon in terms of the number densities of its confined constituents as functions of the longitudinal momentum fraction ($x$) they carry. Our approach, with a truncated Fock space, is appropriate for comparison with experimental results at low energy scales. We then employ QCD evolution of the PDFs to higher momentum scales for comparison with global analyses.

We address the fundamental issue of the contributions of quark and gluon helicity, including sea quarks, to nucleon spin. The RHIC spin program at BNL has shown that gluon helicity ($\Delta G$) is non-zero and likely substantial~\cite{STAR:2014wox,deFlorian:2014yva,Nocera:2014gqa,Ethier:2017zbq}, complementing the known quark helicity contribution of about 30\%. This indicates that parton helicities contribute significantly to nucleon spin. However, there are still large uncertainties regarding the small-$x$ contribution to $\Delta G$, which is defined as the first moment of the polarized gluon PDF~\cite{Ji:2020ena}. Resolving this is a key goal of the future Electron-Ion Colliders (EICs)~\cite{Accardi:2012qut,AbdulKhalek:2021gbh,Anderle:2021wcy}. Addressing various nucleon distribution functions in both longitudinal and transverse momentum space, as well as its spin structure, requires a unified framework. Here, we demonstrate such a framework that successfully captures the qualitative nature of nucleon properties at the hadronic scale.

{\it  LFQCD Hamiltonian and nucleon wave functions.}---The structural information of the nucleon is encoded in the light-front wavefunctions (LFWFs), obtained by solving the Hamiltonian eigenvalue problem: $P^-P^+|{\Psi}\rangle=M^2|{\Psi}\rangle$, where $P^\pm=P^0 \pm P^3$ define the longitudinal momentum ($P^+$) and the light-front Hamiltonian ($P^-$) of the system, with $M^2$ being the mass squared eigenvalue. At fixed LF time,  the nucleon state is expressed as
\begin{align}\label{eq:Fock_space}
|\Psi\rangle=&\psi^{(3q)}|qqq\rangle+\psi^{(3q+g)}|qqqg\rangle + \psi^{(3q+u\bar{u})}|qqqu\bar{u}\rangle\nonumber\\& + \psi^{(3q+d\bar{d})}|qqqd\bar{d}\rangle + \psi^{(3q+s\bar{s})}|qqqs\bar{s}\rangle + \dots\, , 
\end{align}
where $\psi^{(\dots)}$ represents the probability amplitudes for various parton configurations within the nucleon. These amplitudes can define the LFWFs in either coordinate or momentum space.

The LFQCD Hamiltonian that incorporates interactions relevant to those  Fock components in Eq.~\eqref{eq:Fock_space}, in LF gauge, is given by~\cite{Brodsky:1997de}
\begin{align}\label{eqn:PQCD}
P^-_{\rm QCD}=&\int {\rm d}^2x^\perp {\rm d}x^-\Big\{\frac{1}{2}\bar{\Phi}\gamma^+\frac{(m_q+\delta m_q)^2+(i\partial^\perp)^2}{i\partial^+}\Phi\nonumber\\
&-\frac{1}{2}A_a^i[\delta m_{g}^2+(i\partial^\perp)^2]A_a^i+g_s\bar{\Phi}\gamma_\mu T^aA_a^\mu\Phi\nonumber\\
&+\frac{1}{2}g_s^2\bar{\Phi}\gamma^+T^a\Phi\frac{1}{(i\partial^+)^2}\bar{\Phi}\gamma^+T^a\Phi\nonumber\\
&+\frac{g_s^2C_F}{2}\bar{\Phi}\gamma_\mu A^\mu\frac{\gamma^+}{i\partial^+}\gamma_\nu A^\nu\Phi\Big\},
\end{align}
where $\Phi$ and $A^\mu$ denote the quark and gluon fields, respectively. The variables $x^-$ and $x^{\perp}$ define the longitudinal and transverse position coordinates, respectively. $T$ represents the generator of the $SU(3)$ gauge group in color space, and $\gamma^\mu$ are the Dirac matrices. In Eq.~\eqref{eqn:PQCD}, the first and second terms correspond to the kinetic energies of quarks with physical mass $m_q$ and gluons with zero mass, respectively. The last three terms involve vertex and instantaneous interactions with a global coupling constant $g_s$.
A Fock sector-dependent renormalization procedure, initially developed for positronium in the $|e\bar{e}\rangle$ and $|e\bar{e}\gamma\rangle$ bases \cite{Zhao:2014hpa,Zhao:2020kuf}, and subsequently applied to hadrons \cite{Lan:2021wok,Xu:2023nqv}, is used to derive quark and gluon mass counter terms, $\delta m_q$ and $\delta m_g$, respectively. We introduce a distinct quark mass $m_f$ to account for nonperturbative effects in vertex interactions \cite{Burkardt:1998dd,Glazek:1992aq}.

We use the BLFQ framework \cite{Vary:2009gt} to solve the Hamiltonian eigenvalue problem, which has successfully characterized the structure of light mesons \cite{Lan:2021wok} and nucleons \cite{Xu:2023nqv} by incorporating dynamical gluons. Previous studies \cite{Lan:2021wok,Xu:2023nqv} employed an effective confinement interaction and a substantial gluon mass to simulate confinement effects. In contrast, our approach excludes the effective confining potential and adopts the physical gluon mass.
We use a plane-wave state in the longitudinal direction confined in a one-dimensional box of length $2L$, with antiperiodic (periodic) boundary conditions for quarks (gluons). Transversely, we employ a two-dimensional harmonic oscillator (``2D-HO") wave function  $\Phi_{nm}(\vec{p}_\perp;b)$  with scale parameter $b$, and include a light-cone helicity state in spin space \cite{Zhao:2014xaa}. This basis transforms the eigenvalue problem of the Hamiltonian into a matrix eigenvalue problem. Each parton single-particle state in 3D space and spin space is characterized by four quantum numbers  $\bar{\alpha} = \{k, n, m, \lambda\}$.
Here, $k$ represents the longitudinal degree of freedom corresponding to the parton longitudinal momentum $p^{+} = \frac{2\pi k}{L}$, where $k$ takes positive half-integer (integer) values for quarks (gluons); we omit the zero mode for gluons. The 2D-HO wave function is defined by principal quantum number $n$, orbital angular momentum quantum number $m$, and spin $\lambda$. Fock sectors with multiple color-singlet states require additional labels to identify each state. Notably,  $|qqqg\rangle$  and $|qqqq\bar{q}\rangle$ have two and three color-singlet states, respectively.

Two basis space truncations, $N_{\rm max}$ and $K$, are implemented to make the resulting matrix finite \cite{Zhao:2014xaa}. $N_{\rm max}$ truncates the total energy of the 2D-HO basis states in the transverse direction: $\sum_i \left( 2 n_i + |m_i| + 1 \right) \leq N_{\rm max}$. $K$ constrains the sum of longitudinal momenta: $\sum_i k_i = K$, where $x_i = k_i / K$ denotes the longitudinal momentum fraction. Thus, $K$ determines the longitudinal resolution and, consequently, the resolution for the PDFs.

The resulting nucleon LFWFs with helicity $\Lambda$ in momentum space are expressed as components within each Fock sector:
\begin{equation}
\Psi^{\mathcal{N},\,\Lambda}_{\{x_i,\vec{p}_{\perp i},\lambda_i\}} =\sum_{ \{n_i m_i\} }\psi^{\mathcal{N}}({\{\overline{\alpha}_i}\})\prod_{i=1}^{\mathcal{N}}  \phi_{n_i m_i}(\vec{p}_{\perp i},b)\,.
\label{eqn:wf}
\end{equation}
with $\psi^{\mathcal{N}}(\{\overline{\alpha}_i\})$, where $\mathcal{N}$ represents the particle number in each of the Fock sectors $|qqq\rangle$, $|qqqg\rangle$, and $|qqqq\bar{q}\rangle$. 

{\it  Numerical results.}--- All calculations employ $N_{\rm max} = 7$ and $K = 16.5$. The harmonic oscillator scale parameter is set to $b = 0.6$ GeV, while the UV cutoff for the instantaneous interaction is $b_{\rm inst} = 2.80 \pm 0.15$ GeV. Model parameters $\{m_u, m_d, m_f, g_s\} = \{1.0, 0.85, 5.45\pm 0.4, 2.90\pm 0.1\}$ (all in GeV except $g_s$) are determined by fitting the proton mass and electromagnetic properties. Note that the large constituent quark masses in the first two Fock sectors partially account for confinement effects and contribute to the nucleon mass in a QCD bound state with significant binding energy. Note that we have a $0.15$ GeV mass difference between up and down quarks.
In  $|qqqq\bar{q}\rangle$, the partons are treated as free particles due to our neglect of higher Fock sectors. Hence, we use the current quark masses ($m_u=0.0022$ GeV,$m_d=0.0047$ GeV, $m_s=0.094$ GeV) in this Fock sector rather than the model quark masses. At the model scale, the probabilities for finding the proton in a given Fock sector are $53.10\%$ in $|qqq\rangle$, $26.53\%$ in $|qqqg\rangle$, $8.52\%$ in $|qqqu\bar{u}\rangle$, $8.56\%$ in $|qqqd\bar{d}\rangle$, and $3.29\%$ in $|qqqs\bar{s}\rangle$. 

With our resulting LFWFs, the flavor Dirac $F^q_1(Q^2)$ and Pauli $F^q_2(Q^2)$ FFs in the proton can be expressed in terms of overlap integrals as~\cite{Brodsky:2000xy}
\begin{align}\label{eq_DF}
F_1^q(Q^2)=& 
\frac{1}{2}\int_{\mathcal{N}}  \Psi^{\mathcal{N},\,\Lambda\,*}_{\{x_i^\prime,\vec{p}_{\perp i}^{\,\prime},\lambda_i\}}\,\Psi^{\mathcal{N},\,\Lambda}_{\{x_i,\vec{p}_{\perp i},\lambda_i\}},    \\
F_2^q(Q^2)=& -\frac{M }{(q^1-iq^2)}
\int_{\mathcal{N}}  \Psi^{\mathcal{N},\,\Lambda\,*}_{\{x_i^\prime,\vec{p}_{\perp i}^{\,\prime},\lambda_i\}}\,\Psi^{\mathcal{N},\,-\Lambda}_{\{x_i,\vec{p}_{\perp i},\lambda_i\}},
\end{align}
where
$\int_{\mathcal{N}} \equiv  \sum_{\mathcal{N},\,\Lambda,\,\lambda_i}\prod_{i=1}^\mathcal{N} \int \left[\frac{{\rm d}x\,{\rm d}^2\vec{p}_\perp}{16\pi^3}\right]_i16\pi^3$ $  \delta(1-\sum x_j)$ $\delta^2(\sum \vec{p}_{\perp j})$.
We consider the frame where the momentum transfer $q=(0,0,\vec{q}_{\perp})$, thus $Q^2=-q^2=\vec{q}_{\perp}^{\,2}$.
For the struck quark of flavor $q$,
 ${x^\prime}_1=x_1$; ${\vec{p}}_{\perp 1}^{\,\prime}=\vec{p}_{\perp 1}+(1-x_1) \vec{q}_\perp$ and  ${x^\prime}_i={x_i}$; ${\vec{p}}_{\perp i}^{\,\prime}=\vec{p}_{\perp i}-{x_i} \vec{q}_\perp$ for  the spectators.
The nucleon FFs are obtained from the flavor form factors~\cite{Cates:2011pz}. We evaluate the proton Sach's FFs, which are expressed as
\begin{align}
G_{\rm E}(Q^2)=&F_1(Q^2) - \frac{Q^2}{4M^2} F_2(Q^2)\,, \\
G_{\rm M}(Q^2)=&F_1(Q^2) + F_2(Q^2)\,.
\end{align}

\begin{figure}
\begin{center}
\includegraphics[width=0.95\linewidth]{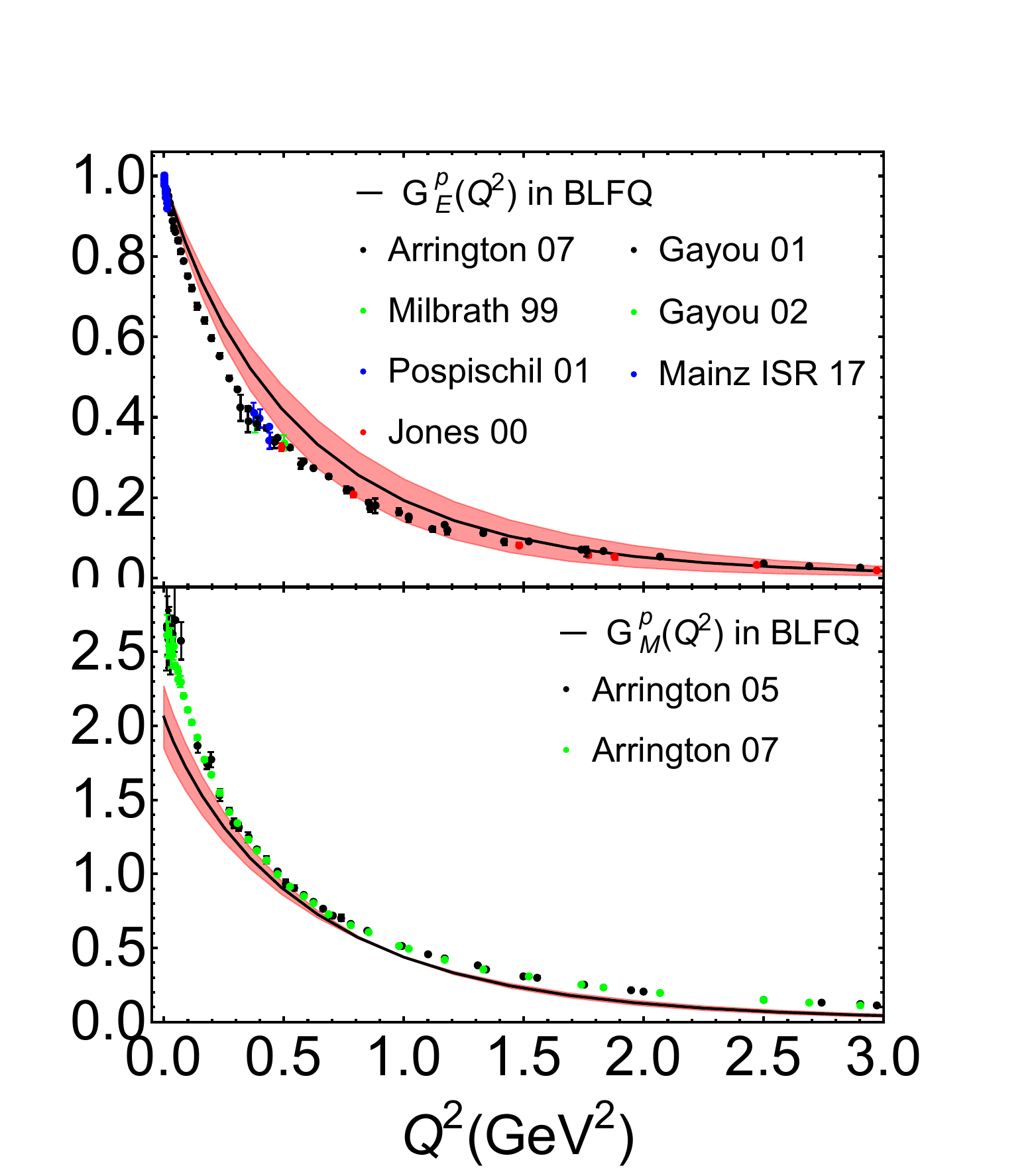}
\caption{The electromagnetic FFs for the proton. Our results (gray bands) are compared with the experimental data taken from Refs.~\cite{Gayou:2001qt,JeffersonLabHallA:1999epl,Arrington:2007ux,JeffersonLabHallA:2001qqe,A1:2001xxy,BatesFPP:1997rpw} for $G^{p}_{\rm E}$ and  Refs.~\cite{Arrington:2004ae,Arrington:2007ux} for $G^{p}_{\rm M}$.}
\label{proton_FFs}
\end{center}
\end{figure}
The electromagnetic FFs of the proton are shown in Fig.~\ref{proton_FFs}.
The red bands represents the uncertainty in the approach due to the uncertainty of the model parameters quoted above.
Overall, our approach shows reasonable agreement with experimental data for the proton electric FF. The proton magnetic FF also aligns reasonably well with data at large  $Q^2$, though it exhibits a 25$\%$ deviation at low $Q^2$.
Electromagnetic radii are computed from the slope of the Sachs form factors~\cite{Ernst:1960zza}, yielding a charge radius $\sqrt{\langle r^2_{E}\rangle} = 0.72\pm 0.05$ fm and a magnetic radius $\sqrt{\langle r^2_{M}\rangle} = 0.73\pm 0.02$ fm, compared to experimental values $\sqrt{\langle r^2_{E}\rangle}_{\rm exp} = 0.840^{+0.003}_{-0.002}$ fm and $\sqrt{\langle r^2_{M}\rangle}_{\rm exp} = 0.849^{+0.003}_{-0.003}$ fm~\cite{Lin:2021xrc,ParticleDataGroup:2024cfk}. 

With our obtained LFWFs, the nucleon's unpolarized, helicity, and helicity PDFs are given by
\begin{align}
&f(x)= \int_{\mathcal{N}}  \frac{1}{2}\,\Psi^{\mathcal{N},\,\Lambda\,*}_{\{x_i,\vec{p}_{\perp i},\lambda_i\}}\,\Psi^{\mathcal{N},\,\Lambda}_{\{x_i,\vec{p}_{\perp i},\lambda_i\}}\,\delta(x-x_i)\,,\\
&\Delta f(x)=  \int_{\mathcal{N}} \frac{\lambda_1}{2}\, \Psi^{\mathcal{N},\,\Lambda\,*}_{\{x_i,\vec{p}_{\perp i},\lambda_i\}}\,\Psi^{\mathcal{N},\,\Lambda}_{\{x_i,\vec{p}_{\perp i},\lambda_i\}}\,\delta(x-x_i)\,,\\
&\delta f(x)= \int_{\mathcal{N}}\, \Psi^{\mathcal{N},\,\Lambda\,*}_{\{x_i,\vec{p}_{\perp i},\lambda_i^\prime\}}\,\Psi^{\mathcal{N},\,-\Lambda}_{\{x_i,\vec{p}_{\perp i},\lambda_i\}}\,\delta(x-x_i) \,,
\label{eqn:pdf_i}
\end{align}
respectively,  where $\lambda^\prime_1=-\lambda_1$ for the struck parton and $\lambda^\prime_{i}=\lambda_{i}$ for the spectators ($i\ne 1$).
The PDFs, interpreted as particle number densities, satisfy the normalization condition for valence quarks: $\int_{0}^1f(x){\rm d}x=n_q$, where $n_q=1$ for down quarks and 
$n_q=2$ for up quarks. Including gluon and sea quark PDFs, the second moment of the PDFs adheres to the sum rule: $\int_0^1\sum_ixf^i(x)\,{\rm d}x=1$.
\begin{figure}
\begin{center}
\includegraphics[width=0.95\linewidth]{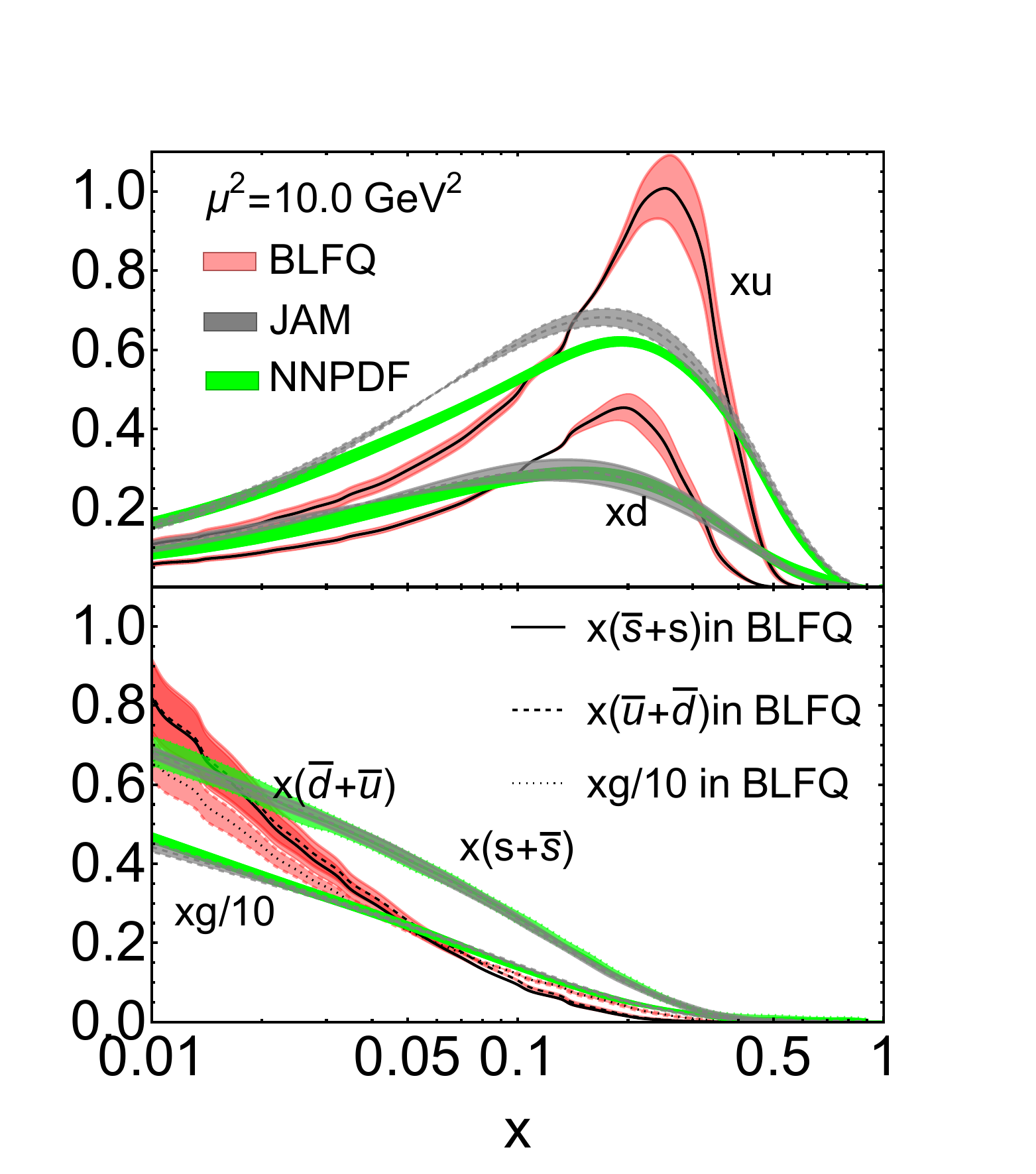}
\caption{The unpolarized PDFs of the proton. Our results (red bands) are compared with the NNPDF3.1~\cite{NNPDF:2017mvq} and JAM~\cite{Cocuzza:2021cbi} global analyses. }
\label{fig_pdf}
\end{center}
\end{figure}
We employ the Higher Order Perturbative Parton Evolution toolkit~\cite{Salam:2008qg}, which numerically solves the Dokshitzer-Gribov-Lipatov-Altarelli-Parisi equations of QCD~~\cite{Dokshitzer:1977sg,Gribov:1972ri,Altarelli:1977zs}, to evolve our PDFs from the model scale  to a higher scale. We determine the model scale $\mu_0^2=0.22\pm 0.02$ GeV$^2$ by matching the second moment of total valence quark PDFs at $\mu^2=10$ GeV$^2$ with the global fitting data, yielding average values $\langle x\rangle_{u+d}=0.37\pm 0.01 $~\cite{NNPDF:2017mvq,Cocuzza:2021cbi}.

Figure~\ref{fig_pdf} illustrates our results for the proton's unpolarized PDFs at  $\mu^2=10$ GeV$^2$, comparing the valence quark and gluon distributions after QCD evolution with NNPDF3.1~\cite{NNPDF:2017mvq} and JAM global analyses~\cite{Cocuzza:2021cbi}. We observe a qualitative agreement between our quark distributions and the global fits. However, due to large constituent quark masses, modeling longitudinal excitations becomes challenging, especially in the absence of the longitudinal confining potential, which significantly influences these excitations. Consequently, our valence quark PDFs are narrower than the global fits in  $x>0.1$ region. The gluon PDF exhibits good consistency with the global fits in the $x>0.05$ region. At small-$x$, where QCD evolution is sensitive to the model scale  $\mu_0$, our gluon PDF qualitatively agrees with the global analyses. 

\begin{figure}
\begin{center}
\includegraphics[width=1.1\linewidth]{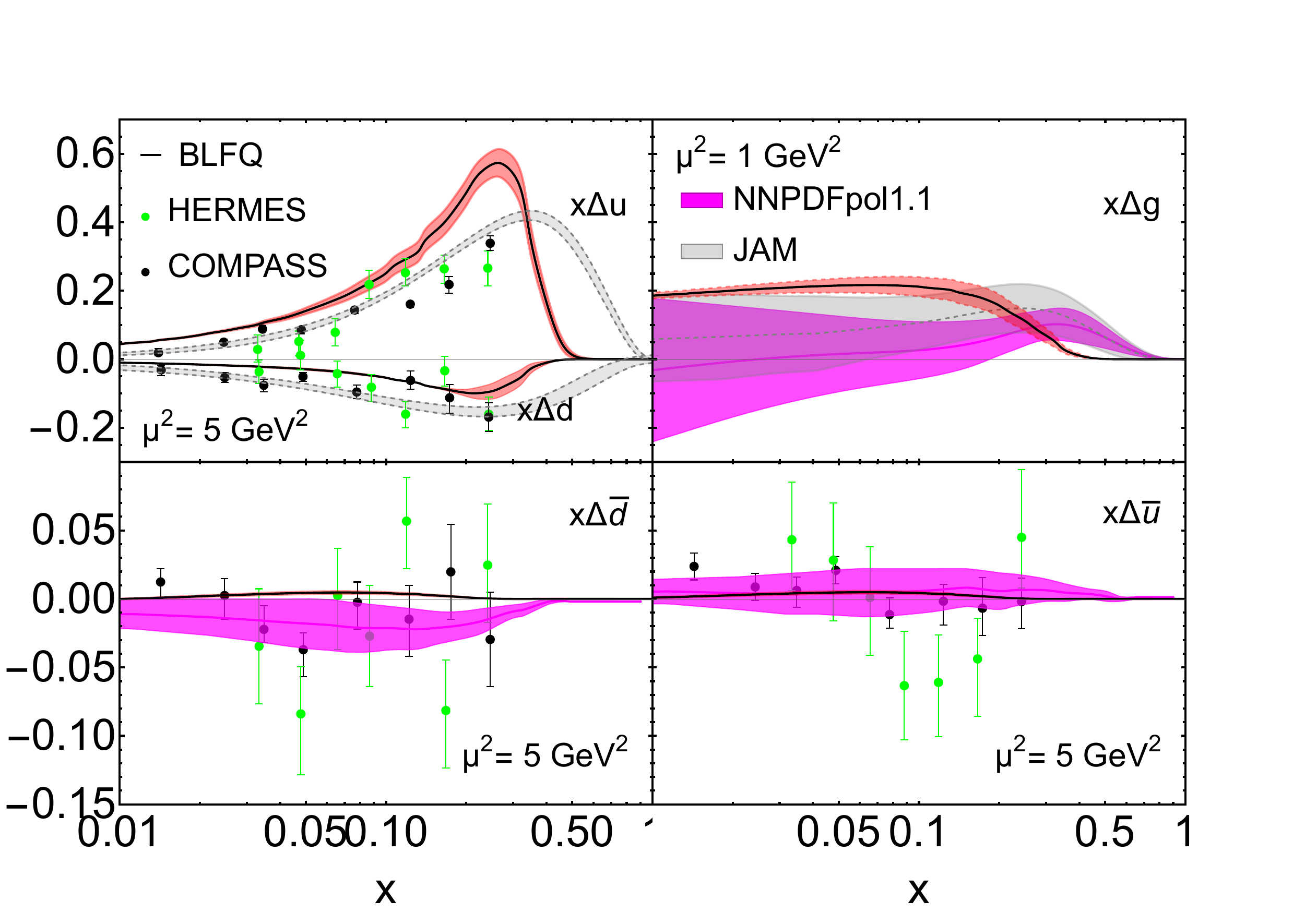}
\caption{The helicity PDFs of the valence $u$ and $d$ (upper-left), $g$ (upper-right), $\bar{u}$ (lower-left), and $\bar{d}$ (lower-right) in the proton. The experimental data are taken from COMPASS~\cite{COMPASS:2010hwr} and HERMES~\cite{HERMES:2003gbu,HERMES:2004zsh} Collaborations. The magenta and gray bands represent the global analysis by NNPDFpol1.1~\cite{Nocera:2014gqa} and JAM~\cite{Sato:2016tuz}, respectively.}
\label{fig_hpdf}
\end{center}
\end{figure}

Figure~\ref{fig_hpdf} displays the helicity PDFs for the valence up and down quarks (upper-left), gluons (upper-right), up-sea quarks (lower-left), and down-sea quarks (lower-right). Our quark helicity PDFs show rough agreement with experimental data from COMPASS~\cite{COMPASS:2010hwr} and HERMES~\cite{HERMES:2003gbu,HERMES:2004zsh}. For comparison, we also include the global analysis by NNPDFpol1.1~\cite{Nocera:2014gqa} and JAM~\cite{Sato:2016tuz}. Similar to the unpolarized PDFs, our valence quark helicity distributions are narrower than the global fits in $x> 0.1$. Note that the signs of the sea quark helicity PDFs are not determined by experimental data.

We present the gluon helicity PDF at the scale $\mu^2=1$ GeV$^2$ and compare our prediction with the global analyses by the JAM~\cite{Sato:2016tuz} and NNPDF Collaborations~\cite{Nocera:2014gqa}. Our prediction shows a rough agreement with the global fits at small-$x$, while our gluon helicity distribution at large-$x$ decreases more rapidly than the global analyses. Our prediction aligns better with the JAM results up to $x \sim 0.2$. Note that significant uncertainties in the global analyses persist, particularly in the small-$x$ region where even the sign is uncertain~\cite{Zhou:2022wzm}.

The partonic helicity contributions to the proton spin are determined by the first moment of the helicity distributions, known as the axial charge. Our analysis shows that, at the model scale, the axial charges of the up and down quarks are $\Delta \Sigma_u=0.89\pm 0.07$ and $\Delta \Sigma_d=-0.22 \pm 0.02$, respectively. Our prediction for $\Delta \Sigma_u$ aligns well with the world data summarized in Ref.~\cite{Deur:2018roz}, while $\Delta \Sigma_d$ is somewhat smaller than most of the world data. Consequently, the quark helicity contribution to the proton spin, $\frac{1}{2}\Delta \Sigma=0.33\pm 0.04$, is somewhat larger compared to the world data. We find a significant gluon contribution, $\Delta G=0.29\pm 0.03$ for $x_g\in [0.05, 0.2]$ at $10$ GeV$^2$, to the proton spin, which agrees well with the NNPDF analysis: $\Delta G=0.23 (6)$~\cite{Nocera:2014gqa} and the lattice QCD prediction at the physical pion mass: $\Delta G = 0.251(47)(16)$~\cite{Yang:2016plb}.  In the light-cone gauge, following the sum rules of GPDs~\cite{Ji:1996ek}: $L_z=\int {\rm d}x\{\frac{1}{2}x[H(x,0,0)+E(x,0,0)]-\Tilde{H}(x,0,0)\}$, we calculate the partonic orbital angular momentum (OAM) contributions to the proton spin. Our numerical results show that $L_z^u=0.036\pm 0.002$, $L_z^d=0.058\pm 0.001$ and $L_z^g=-0.037\pm 0.004$. Currently, nothing is known experimentally about 
OAM. However, gluon OAM can be extracted from double spin asymmetry in diffractive dijet production~\cite{Bhattacharya:2022vvo}, while quark OAM can be measured in the exclusive double Drell-Yan process~\cite{Bhattacharya:2017bvs}. 
\begin{figure}
\begin{center}
\includegraphics[width=0.95\linewidth]{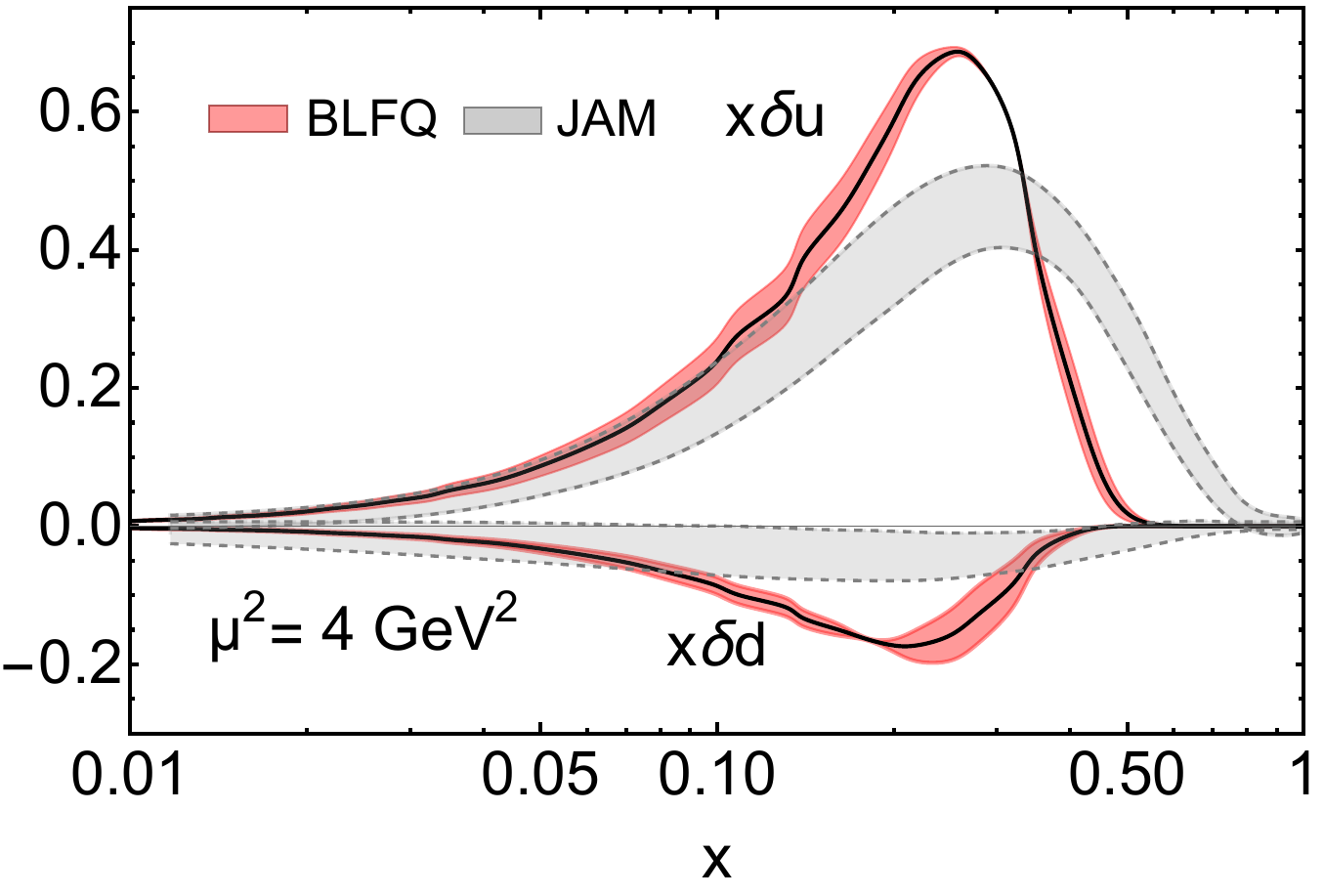}
\caption{Comparison for the transversity PDF in the proton from BLFQ (red
bands) and the recent global analyses by JAM~\cite{Cocuzza:2023oam} (gray band).
}
\label{fig_h1}
\end{center}
\end{figure}

The transversity PDF measures the extent of transverse polarization of quarks within a transversely polarized nucleon. In Fig.~\ref{fig_h1}, we compare our predictions for the transversity PDFs with the recent JAM analysis at $\mu^2=4$ GeV$^2$~\cite{Cocuzza:2023oam}. Our calculations agree with the global fit in the $x<0.1$ region. However, we observe a narrow peak at large $x$ for both up and down quarks, similar to the unpolarized and helicity PDFs. Furthermore, the transversity and helicity PDFs show a symmetric behavior for up and down quarks, a symmetry disrupted by the effective confining potential in our previous study~\cite{Mondal:2019jdg}. We suggest that this symmetry could be altered by including higher Fock sectors and other QCD interactions. Notably, the transversity PDFs of gluons cannot be defined in the same manner as those for quarks.

The tensor charge is obtained from the first moment of the transversity PDFs. At the scale 
$\mu^2=4$ GeV$^2$, our calculations yield a tensor charge of 
$\delta u=0.81\pm 0.08$ for the up quark, which is somewhat larger than the recent JAM analysis 
$\delta u = 0.71 (2)$~\cite{Cocuzza:2023oam}, as well as the lattice QCD result of 
$\delta u = 0.784 (28)$~\cite{Gupta:2018lvp}. For the down quark, our prediction of 
$\delta d=-0.22\pm 0.01$ agrees well with the recently extracted data of 
$\delta d = 0.200(6)$~\cite{Cocuzza:2023oam} and the lattice QCD prediction of 
$\delta d = -0.204 (11)$~\cite{Gupta:2018lvp}. It is important to note that all these observables can, in principle, be systematically improved by incorporating higher Fock components and relevant QCD interactions.

{\it Conclusion and outlooks.}---
%
%
We have solved the LFQCD Hamiltonian for the nucleon within the combined constituent three quarks ($|qqq\rangle$), three quarks and one gluon ($|qqqg\rangle$), and three quarks and a pair of quark-antiquark ($|qqqq\bar{q}\rangle$) Fock spaces using BLFQ, obtaining LFWFs to compute the nucleon's partonic structure. 
We have calculated the electromagnetic FFs and quark and gluon unpolarized, helicity, and transversity PDFs, observing qualitative agreement between our results and the corresponding experimental data or global fits. 

We have calculated the quark and gluon helicity and OAM contributions to the proton spin sum rule. Our predictions indicate that quark helicity contributes $66\%$ and gluon helicity $12\%$ to the proton's total angular momentum. The OAM contributions are $L_z^u=0.036\pm 0.002$, $L_z^d=0.058\pm 0.001$, and $L_z^g=-0.037\pm 0.004$. Large uncertainties remain in gluon helicity, especially in the small-$x$ domain, making future measurements in the $x_g<0.02$ region crucial. Currently, nothing is known experimentally about OAM. Resolving these issues is a major goal for future EICs. Our approach yields tensor charges of $\delta u=0.81\pm 0.08$ and $\delta d=-0.22\pm 0.01$, with the down quark's value aligning well with extracted data and lattice QCD predictions, while the up quark's value is somewhat larger than these benchmarks. 

While our current approach yields somewhat less satisfactory results for the parton distributions, particularly at large-$x$, it remains acceptable as it can be systematically improved. Notably, the inclusion of additional Fock sectors with multi-gluon configurations, relevant for three-gluon and four-gluon interactions, are expected to play an important role in generating color confinement.
Having said that the obtained LFWFs represent a significant advancement towards a unified description of various hadron distribution functions in both longitudinal and transverse momentum space. They can be further utilized to compute quark and gluon GPDs, TMDs, Wigner distributions, and double parton correlations within the nucleon. 

\begin{acknowledgments}
{\it Acknowledgments.}---CM is supported by new faculty start up funding by the Institute of Modern Physics, Chinese Academy of Sciences, Grant No. E129952YR0.  CM also thanks the Chinese Academy of Sciences Presidents International Fellowship Initiative for the support via Grants No. 2021PM0023. JL is supported by Special Research Assistant Funding Project, Chinese Academy of Sciences, by the Natural Science Foundation of Gansu Province, China, Grant No.23JRRA631, and by National Natural Science Foundation of China, Grant No. 12305095. XZ is supported by new faculty startup funding by the Institute of Modern Physics, Chinese Academy of Sciences, by Key Research Program of Frontier Sciences, Chinese Academy of Sciences, Grant No. ZDBS-LY-7020, by the Natural Science Foundation of Gansu Province, China, Grant No. 20JR10RA067, by the Foundation for Key Talents of Gansu Province, by the Central Funds Guiding the Local Science and Technology Development of Gansu Province and by the Strategic Priority Research Program of the Chinese Academy of Sciences, Grant No. XDB34000000. This research is supported by Gansu International Collaboration and Talents Recruitment Base of Particle Physics (2023-2027), and supported by the International Partnership Program of Chinese Academy of Sciences, Grant No.016GJHZ2022103FN. YL is supported by the new faculty startup fund of University of Science and Technology of
China, by the National Natural Science Foundation of China (NSFC) under Grant No. 12375081, by the Chinese Academy of Sciences under Grant
No. YSBR-101. JPV is supported by the U.S. Department of Energy under Grant No. DE-SC0023692.  A portion of the computational resources were also provided by Gansu Computing Center and by Sugon Computing Center in Xi'an. 
\end{acknowledgments}

\bibliography{refs-ent-blfq}

\begin{thebibliography}{64}%
\makeatletter
\providecommand \@ifxundefined [1]{%
 \@ifx{#1\undefined}
}%
\providecommand \@ifnum [1]{%
 \ifnum #1\expandafter \@firstoftwo
 \else \expandafter \@secondoftwo
 \fi
}%
\providecommand \@ifx [1]{%
 \ifx #1\expandafter \@firstoftwo
 \else \expandafter \@secondoftwo
 \fi
}%
\providecommand \natexlab [1]{#1}%
\providecommand \enquote  [1]{``#1''}%
\providecommand \bibnamefont  [1]{#1}%
\providecommand \bibfnamefont [1]{#1}%
\providecommand \citenamefont [1]{#1}%
\providecommand \href@noop [0]{\@secondoftwo}%
\providecommand \href [0]{\begingroup \@sanitize@url \@href}%
\providecommand \@href[1]{\@@startlink{#1}\@@href}%
\providecommand \@@href[1]{\endgroup#1\@@endlink}%
\providecommand \@sanitize@url [0]{\catcode `\\12\catcode `\$12\catcode `\&12\catcode `\#12\catcode `\^12\catcode `\_12\catcode `\%12\relax}%
\providecommand \@@startlink[1]{}%
\providecommand \@@endlink[0]{}%
\providecommand \url  [0]{\begingroup\@sanitize@url \@url }%
\providecommand \@url [1]{\endgroup\@href {#1}{\urlprefix }}%
\providecommand \urlprefix  [0]{URL }%
\providecommand \Eprint [0]{\href }%
\providecommand \doibase [0]{http://dx.doi.org/}%
\providecommand \selectlanguage [0]{\@gobble}%
\providecommand \bibinfo  [0]{\@secondoftwo}%
\providecommand \bibfield  [0]{\@secondoftwo}%
\providecommand \translation [1]{[#1]}%
\providecommand \BibitemOpen [0]{}%
\providecommand \bibitemStop [0]{}%
\providecommand \bibitemNoStop [0]{.\EOS\space}%
\providecommand \EOS [0]{\spacefactor3000\relax}%
\providecommand \BibitemShut  [1]{\csname bibitem#1\endcsname}%
\let\auto@bib@innerbib\@empty
\bibitem [{\citenamefont {Callan}\ \emph {et~al.}(1978)\citenamefont {Callan}, \citenamefont {Dashen},\ and\ \citenamefont {Gross}}]{Callan:1977gz}%
  \BibitemOpen
  \bibfield  {author} {\bibinfo {author} {\bibfnamefont {C.~G.}\ \bibnamefont {Callan}, \bibfnamefont {Jr.}}, \bibinfo {author} {\bibfnamefont {R.~F.}\ \bibnamefont {Dashen}}, \ and\ \bibinfo {author} {\bibfnamefont {D.~J.}\ \bibnamefont {Gross}},\ }\href {\doibase 10.1103/PhysRevD.17.2717} {\bibfield  {journal} {\bibinfo  {journal} {Phys. Rev. D}\ }\textbf {\bibinfo {volume} {17}},\ \bibinfo {pages} {2717} (\bibinfo {year} {1978})}\BibitemShut {NoStop}%
\bibitem [{\citenamefont {Hagler}(2010)}]{Hagler:2009ni}%
  \BibitemOpen
  \bibfield  {author} {\bibinfo {author} {\bibfnamefont {P.}~\bibnamefont {Hagler}},\ }\href {\doibase 10.1016/j.physrep.2009.12.008} {\bibfield  {journal} {\bibinfo  {journal} {Phys. Rept.}\ }\textbf {\bibinfo {volume} {490}},\ \bibinfo {pages} {49} (\bibinfo {year} {2010})},\ \Eprint {http://arxiv.org/abs/0912.5483} {arXiv:0912.5483 [hep-lat]} \BibitemShut {NoStop}%
\bibitem [{\citenamefont {Jo\'o}\ \emph {et~al.}(2019)\citenamefont {Jo\'o}, \citenamefont {Jung}, \citenamefont {Christ}, \citenamefont {Detmold}, \citenamefont {Edwards}, \citenamefont {Savage},\ and\ \citenamefont {Shanahan}}]{Joo:2019byq}%
  \BibitemOpen
  \bibfield  {author} {\bibinfo {author} {\bibfnamefont {B.}~\bibnamefont {Jo\'o}}, \bibinfo {author} {\bibfnamefont {C.}~\bibnamefont {Jung}}, \bibinfo {author} {\bibfnamefont {N.~H.}\ \bibnamefont {Christ}}, \bibinfo {author} {\bibfnamefont {W.}~\bibnamefont {Detmold}}, \bibinfo {author} {\bibfnamefont {R.}~\bibnamefont {Edwards}}, \bibinfo {author} {\bibfnamefont {M.}~\bibnamefont {Savage}}, \ and\ \bibinfo {author} {\bibfnamefont {P.}~\bibnamefont {Shanahan}} (\bibinfo {collaboration} {USQCD}),\ }\href {\doibase 10.1140/epja/i2019-12919-7} {\bibfield  {journal} {\bibinfo  {journal} {Eur. Phys. J. A}\ }\textbf {\bibinfo {volume} {55}},\ \bibinfo {pages} {199} (\bibinfo {year} {2019})},\ \Eprint {http://arxiv.org/abs/1904.09725} {arXiv:1904.09725 [hep-lat]} \BibitemShut {NoStop}%
\bibitem [{\citenamefont {Bazavov}\ \emph {et~al.}(2010)\citenamefont {Bazavov} \emph {et~al.}}]{MILC:2009mpl}%
  \BibitemOpen
  \bibfield  {author} {\bibinfo {author} {\bibfnamefont {A.}~\bibnamefont {Bazavov}} \emph {et~al.} (\bibinfo {collaboration} {MILC}),\ }\href {\doibase 10.1103/RevModPhys.82.1349} {\bibfield  {journal} {\bibinfo  {journal} {Rev. Mod. Phys.}\ }\textbf {\bibinfo {volume} {82}},\ \bibinfo {pages} {1349} (\bibinfo {year} {2010})},\ \Eprint {http://arxiv.org/abs/0903.3598} {arXiv:0903.3598 [hep-lat]} \BibitemShut {NoStop}%
\bibitem [{\citenamefont {Durr}\ \emph {et~al.}(2008)\citenamefont {Durr} \emph {et~al.}}]{BMW:2008jgk}%
  \BibitemOpen
  \bibfield  {author} {\bibinfo {author} {\bibfnamefont {S.}~\bibnamefont {Durr}} \emph {et~al.} (\bibinfo {collaboration} {BMW}),\ }\href {\doibase 10.1126/science.1163233} {\bibfield  {journal} {\bibinfo  {journal} {Science}\ }\textbf {\bibinfo {volume} {322}},\ \bibinfo {pages} {1224} (\bibinfo {year} {2008})},\ \Eprint {http://arxiv.org/abs/0906.3599} {arXiv:0906.3599 [hep-lat]} \BibitemShut {NoStop}%
\bibitem [{\citenamefont {Maris}\ and\ \citenamefont {Roberts}(2003)}]{Maris:2003vk}%
  \BibitemOpen
  \bibfield  {author} {\bibinfo {author} {\bibfnamefont {P.}~\bibnamefont {Maris}}\ and\ \bibinfo {author} {\bibfnamefont {C.~D.}\ \bibnamefont {Roberts}},\ }\href {\doibase 10.1142/S0218301303001326} {\bibfield  {journal} {\bibinfo  {journal} {Int. J. Mod. Phys. E}\ }\textbf {\bibinfo {volume} {12}},\ \bibinfo {pages} {297} (\bibinfo {year} {2003})},\ \Eprint {http://arxiv.org/abs/nucl-th/0301049} {arXiv:nucl-th/0301049} \BibitemShut {NoStop}%
\bibitem [{\citenamefont {Roberts}\ and\ \citenamefont {Williams}(1994)}]{Roberts:1994dr}%
  \BibitemOpen
  \bibfield  {author} {\bibinfo {author} {\bibfnamefont {C.~D.}\ \bibnamefont {Roberts}}\ and\ \bibinfo {author} {\bibfnamefont {A.~G.}\ \bibnamefont {Williams}},\ }\href {\doibase 10.1016/0146-6410(94)90049-3} {\bibfield  {journal} {\bibinfo  {journal} {Prog. Part. Nucl. Phys.}\ }\textbf {\bibinfo {volume} {33}},\ \bibinfo {pages} {477} (\bibinfo {year} {1994})},\ \Eprint {http://arxiv.org/abs/hep-ph/9403224} {arXiv:hep-ph/9403224} \BibitemShut {NoStop}%
\bibitem [{\citenamefont {Bashir}\ \emph {et~al.}(2012)\citenamefont {Bashir}, \citenamefont {Chang}, \citenamefont {Cloet}, \citenamefont {El-Bennich}, \citenamefont {Liu}, \citenamefont {Roberts},\ and\ \citenamefont {Tandy}}]{Bashir:2012fs}%
  \BibitemOpen
  \bibfield  {author} {\bibinfo {author} {\bibfnamefont {A.}~\bibnamefont {Bashir}}, \bibinfo {author} {\bibfnamefont {L.}~\bibnamefont {Chang}}, \bibinfo {author} {\bibfnamefont {I.~C.}\ \bibnamefont {Cloet}}, \bibinfo {author} {\bibfnamefont {B.}~\bibnamefont {El-Bennich}}, \bibinfo {author} {\bibfnamefont {Y.-X.}\ \bibnamefont {Liu}}, \bibinfo {author} {\bibfnamefont {C.~D.}\ \bibnamefont {Roberts}}, \ and\ \bibinfo {author} {\bibfnamefont {P.~C.}\ \bibnamefont {Tandy}},\ }\href {\doibase 10.1088/0253-6102/58/1/16} {\bibfield  {journal} {\bibinfo  {journal} {Commun. Theor. Phys.}\ }\textbf {\bibinfo {volume} {58}},\ \bibinfo {pages} {79} (\bibinfo {year} {2012})},\ \Eprint {http://arxiv.org/abs/1201.3366} {arXiv:1201.3366 [nucl-th]} \BibitemShut {NoStop}%
\bibitem [{\citenamefont {Brodsky}\ \emph {et~al.}(1998)\citenamefont {Brodsky}, \citenamefont {Pauli},\ and\ \citenamefont {Pinsky}}]{Brodsky:1997de}%
  \BibitemOpen
  \bibfield  {author} {\bibinfo {author} {\bibfnamefont {S.~J.}\ \bibnamefont {Brodsky}}, \bibinfo {author} {\bibfnamefont {H.-C.}\ \bibnamefont {Pauli}}, \ and\ \bibinfo {author} {\bibfnamefont {S.~S.}\ \bibnamefont {Pinsky}},\ }\href {\doibase 10.1016/S0370-1573(97)00089-6} {\bibfield  {journal} {\bibinfo  {journal} {Physics Reports}\ }\textbf {\bibinfo {volume} {301}},\ \bibinfo {pages} {299} (\bibinfo {year} {1998})},\ \Eprint {http://arxiv.org/abs/hep-ph/9705477} {arXiv:hep-ph/9705477} \BibitemShut {NoStop}%
\bibitem [{\citenamefont {Brodsky}\ \emph {et~al.}(2015)\citenamefont {Brodsky}, \citenamefont {de~Teramond}, \citenamefont {Dosch},\ and\ \citenamefont {Erlich}}]{Brodsky:2014yha}%
  \BibitemOpen
  \bibfield  {author} {\bibinfo {author} {\bibfnamefont {S.~J.}\ \bibnamefont {Brodsky}}, \bibinfo {author} {\bibfnamefont {G.~F.}\ \bibnamefont {de~Teramond}}, \bibinfo {author} {\bibfnamefont {H.~G.}\ \bibnamefont {Dosch}}, \ and\ \bibinfo {author} {\bibfnamefont {J.}~\bibnamefont {Erlich}},\ }\href {\doibase 10.1016/j.physrep.2015.05.001} {\bibfield  {journal} {\bibinfo  {journal} {Phys. Rept.}\ }\textbf {\bibinfo {volume} {584}},\ \bibinfo {pages} {1} (\bibinfo {year} {2015})},\ \Eprint {http://arxiv.org/abs/1407.8131} {arXiv:1407.8131 [hep-ph]} \BibitemShut {NoStop}%
\bibitem [{\citenamefont {Vary}\ \emph {et~al.}(2010)\citenamefont {Vary}, \citenamefont {Honkanen}, \citenamefont {Li}, \citenamefont {Maris}, \citenamefont {Brodsky}, \citenamefont {Harindranath}, \citenamefont {de~Teramond}, \citenamefont {Sternberg}, \citenamefont {Ng},\ and\ \citenamefont {Yang}}]{Vary:2009gt}%
  \BibitemOpen
  \bibfield  {author} {\bibinfo {author} {\bibfnamefont {J.~P.}\ \bibnamefont {Vary}}, \bibinfo {author} {\bibfnamefont {H.}~\bibnamefont {Honkanen}}, \bibinfo {author} {\bibfnamefont {J.}~\bibnamefont {Li}}, \bibinfo {author} {\bibfnamefont {P.}~\bibnamefont {Maris}}, \bibinfo {author} {\bibfnamefont {S.~J.}\ \bibnamefont {Brodsky}}, \bibinfo {author} {\bibfnamefont {A.}~\bibnamefont {Harindranath}}, \bibinfo {author} {\bibfnamefont {G.~F.}\ \bibnamefont {de~Teramond}}, \bibinfo {author} {\bibfnamefont {P.}~\bibnamefont {Sternberg}}, \bibinfo {author} {\bibfnamefont {E.~G.}\ \bibnamefont {Ng}}, \ and\ \bibinfo {author} {\bibfnamefont {C.}~\bibnamefont {Yang}},\ }\href {\doibase 10.1103/PhysRevC.81.035205} {\bibfield  {journal} {\bibinfo  {journal} {Phys. Rev. C}\ }\textbf {\bibinfo {volume} {81}},\ \bibinfo {pages} {035205} (\bibinfo {year} {2010})},\ \Eprint {http://arxiv.org/abs/0905.1411} {arXiv:0905.1411 [nucl-th]} \BibitemShut {NoStop}%
\bibitem [{\citenamefont {Zhao}\ \emph {et~al.}(2014)\citenamefont {Zhao}, \citenamefont {Honkanen}, \citenamefont {Maris}, \citenamefont {Vary},\ and\ \citenamefont {Brodsky}}]{Zhao:2014xaa}%
  \BibitemOpen
  \bibfield  {author} {\bibinfo {author} {\bibfnamefont {X.}~\bibnamefont {Zhao}}, \bibinfo {author} {\bibfnamefont {H.}~\bibnamefont {Honkanen}}, \bibinfo {author} {\bibfnamefont {P.}~\bibnamefont {Maris}}, \bibinfo {author} {\bibfnamefont {J.~P.}\ \bibnamefont {Vary}}, \ and\ \bibinfo {author} {\bibfnamefont {S.~J.}\ \bibnamefont {Brodsky}},\ }\href {\doibase 10.1016/j.physletb.2014.08.020} {\bibfield  {journal} {\bibinfo  {journal} {Phys. Lett. B}\ }\textbf {\bibinfo {volume} {737}},\ \bibinfo {pages} {65} (\bibinfo {year} {2014})},\ \Eprint {http://arxiv.org/abs/1402.4195} {arXiv:1402.4195 [nucl-th]} \BibitemShut {NoStop}%
\bibitem [{\citenamefont {Nair}\ \emph {et~al.}(2022)\citenamefont {Nair}, \citenamefont {Mondal}, \citenamefont {Zhao}, \citenamefont {Mukherjee},\ and\ \citenamefont {Vary}}]{Nair:2022evk}%
  \BibitemOpen
  \bibfield  {author} {\bibinfo {author} {\bibfnamefont {S.}~\bibnamefont {Nair}}, \bibinfo {author} {\bibfnamefont {C.}~\bibnamefont {Mondal}}, \bibinfo {author} {\bibfnamefont {X.}~\bibnamefont {Zhao}}, \bibinfo {author} {\bibfnamefont {A.}~\bibnamefont {Mukherjee}}, \ and\ \bibinfo {author} {\bibfnamefont {J.~P.}\ \bibnamefont {Vary}} (\bibinfo {collaboration} {BLFQ}),\ }\href {\doibase 10.1016/j.physletb.2022.137005} {\bibfield  {journal} {\bibinfo  {journal} {Phys. Lett. B}\ }\textbf {\bibinfo {volume} {827}},\ \bibinfo {pages} {137005} (\bibinfo {year} {2022})},\ \Eprint {http://arxiv.org/abs/2201.12770} {arXiv:2201.12770 [hep-ph]} \BibitemShut {NoStop}%
\bibitem [{\citenamefont {Wiecki}\ \emph {et~al.}(2015)\citenamefont {Wiecki}, \citenamefont {Li}, \citenamefont {Zhao}, \citenamefont {Maris},\ and\ \citenamefont {Vary}}]{Wiecki:2014ola}%
  \BibitemOpen
  \bibfield  {author} {\bibinfo {author} {\bibfnamefont {P.}~\bibnamefont {Wiecki}}, \bibinfo {author} {\bibfnamefont {Y.}~\bibnamefont {Li}}, \bibinfo {author} {\bibfnamefont {X.}~\bibnamefont {Zhao}}, \bibinfo {author} {\bibfnamefont {P.}~\bibnamefont {Maris}}, \ and\ \bibinfo {author} {\bibfnamefont {J.~P.}\ \bibnamefont {Vary}},\ }\href {\doibase 10.1103/PhysRevD.91.105009} {\bibfield  {journal} {\bibinfo  {journal} {Phys. Rev. D}\ }\textbf {\bibinfo {volume} {91}},\ \bibinfo {pages} {105009} (\bibinfo {year} {2015})},\ \Eprint {http://arxiv.org/abs/1404.6234} {arXiv:1404.6234 [nucl-th]} \BibitemShut {NoStop}%
\bibitem [{\citenamefont {Li}\ \emph {et~al.}(2016)\citenamefont {Li}, \citenamefont {Maris}, \citenamefont {Zhao},\ and\ \citenamefont {Vary}}]{Li:2015zda}%
  \BibitemOpen
  \bibfield  {author} {\bibinfo {author} {\bibfnamefont {Y.}~\bibnamefont {Li}}, \bibinfo {author} {\bibfnamefont {P.}~\bibnamefont {Maris}}, \bibinfo {author} {\bibfnamefont {X.}~\bibnamefont {Zhao}}, \ and\ \bibinfo {author} {\bibfnamefont {J.~P.}\ \bibnamefont {Vary}},\ }\href {\doibase 10.1016/j.physletb.2016.04.065} {\bibfield  {journal} {\bibinfo  {journal} {Phys. Lett. B}\ }\textbf {\bibinfo {volume} {758}},\ \bibinfo {pages} {118} (\bibinfo {year} {2016})},\ \Eprint {http://arxiv.org/abs/1509.07212} {arXiv:1509.07212 [hep-ph]} \BibitemShut {NoStop}%
\bibitem [{\citenamefont {Jia}\ and\ \citenamefont {Vary}(2019)}]{Jia:2018ary}%
  \BibitemOpen
  \bibfield  {author} {\bibinfo {author} {\bibfnamefont {S.}~\bibnamefont {Jia}}\ and\ \bibinfo {author} {\bibfnamefont {J.~P.}\ \bibnamefont {Vary}},\ }\href {\doibase 10.1103/PhysRevC.99.035206} {\bibfield  {journal} {\bibinfo  {journal} {Phys. Rev. C}\ }\textbf {\bibinfo {volume} {99}},\ \bibinfo {pages} {035206} (\bibinfo {year} {2019})},\ \Eprint {http://arxiv.org/abs/1811.08512} {arXiv:1811.08512 [nucl-th]} \BibitemShut {NoStop}%
\bibitem [{\citenamefont {Lan}\ \emph {et~al.}(2019)\citenamefont {Lan}, \citenamefont {Mondal}, \citenamefont {Jia}, \citenamefont {Zhao},\ and\ \citenamefont {Vary}}]{Lan:2019vui}%
  \BibitemOpen
  \bibfield  {author} {\bibinfo {author} {\bibfnamefont {J.}~\bibnamefont {Lan}}, \bibinfo {author} {\bibfnamefont {C.}~\bibnamefont {Mondal}}, \bibinfo {author} {\bibfnamefont {S.}~\bibnamefont {Jia}}, \bibinfo {author} {\bibfnamefont {X.}~\bibnamefont {Zhao}}, \ and\ \bibinfo {author} {\bibfnamefont {J.~P.}\ \bibnamefont {Vary}},\ }\href {\doibase 10.1103/PhysRevLett.122.172001} {\bibfield  {journal} {\bibinfo  {journal} {Phys. Rev. Lett.}\ }\textbf {\bibinfo {volume} {122}},\ \bibinfo {pages} {172001} (\bibinfo {year} {2019})},\ \Eprint {http://arxiv.org/abs/1901.11430} {arXiv:1901.11430 [nucl-th]} \BibitemShut {NoStop}%
\bibitem [{\citenamefont {Mondal}\ \emph {et~al.}(2020)\citenamefont {Mondal}, \citenamefont {Xu}, \citenamefont {Lan}, \citenamefont {Zhao}, \citenamefont {Li}, \citenamefont {Chakrabarti},\ and\ \citenamefont {Vary}}]{Mondal:2019jdg}%
  \BibitemOpen
  \bibfield  {author} {\bibinfo {author} {\bibfnamefont {C.}~\bibnamefont {Mondal}}, \bibinfo {author} {\bibfnamefont {S.}~\bibnamefont {Xu}}, \bibinfo {author} {\bibfnamefont {J.}~\bibnamefont {Lan}}, \bibinfo {author} {\bibfnamefont {X.}~\bibnamefont {Zhao}}, \bibinfo {author} {\bibfnamefont {Y.}~\bibnamefont {Li}}, \bibinfo {author} {\bibfnamefont {D.}~\bibnamefont {Chakrabarti}}, \ and\ \bibinfo {author} {\bibfnamefont {J.~P.}\ \bibnamefont {Vary}},\ }\href {\doibase 10.1103/PhysRevD.102.016008} {\bibfield  {journal} {\bibinfo  {journal} {Phys. Rev. D}\ }\textbf {\bibinfo {volume} {102}},\ \bibinfo {pages} {016008} (\bibinfo {year} {2020})},\ \Eprint {http://arxiv.org/abs/1911.10913} {arXiv:1911.10913 [hep-ph]} \BibitemShut {NoStop}%
\bibitem [{\citenamefont {Xu}\ \emph {et~al.}(2021)\citenamefont {Xu}, \citenamefont {Mondal}, \citenamefont {Lan}, \citenamefont {Zhao}, \citenamefont {Li},\ and\ \citenamefont {Vary}}]{Xu:2021wwj}%
  \BibitemOpen
  \bibfield  {author} {\bibinfo {author} {\bibfnamefont {S.}~\bibnamefont {Xu}}, \bibinfo {author} {\bibfnamefont {C.}~\bibnamefont {Mondal}}, \bibinfo {author} {\bibfnamefont {J.}~\bibnamefont {Lan}}, \bibinfo {author} {\bibfnamefont {X.}~\bibnamefont {Zhao}}, \bibinfo {author} {\bibfnamefont {Y.}~\bibnamefont {Li}}, \ and\ \bibinfo {author} {\bibfnamefont {J.~P.}\ \bibnamefont {Vary}} (\bibinfo {collaboration} {BLFQ}),\ }\href {\doibase 10.1103/PhysRevD.104.094036} {\bibfield  {journal} {\bibinfo  {journal} {Phys. Rev. D}\ }\textbf {\bibinfo {volume} {104}},\ \bibinfo {pages} {094036} (\bibinfo {year} {2021})},\ \Eprint {http://arxiv.org/abs/2108.03909} {arXiv:2108.03909 [hep-ph]} \BibitemShut {NoStop}%
\bibitem [{\citenamefont {Kuang}\ \emph {et~al.}(2022)\citenamefont {Kuang}, \citenamefont {Serafin}, \citenamefont {Zhao},\ and\ \citenamefont {Vary}}]{Kuang:2022vdy}%
  \BibitemOpen
  \bibfield  {author} {\bibinfo {author} {\bibfnamefont {Z.}~\bibnamefont {Kuang}}, \bibinfo {author} {\bibfnamefont {K.}~\bibnamefont {Serafin}}, \bibinfo {author} {\bibfnamefont {X.}~\bibnamefont {Zhao}}, \ and\ \bibinfo {author} {\bibfnamefont {J.~P.}\ \bibnamefont {Vary}} (\bibinfo {collaboration} {BLFQ}),\ }\href {\doibase 10.1103/PhysRevD.105.094028} {\bibfield  {journal} {\bibinfo  {journal} {Phys. Rev. D}\ }\textbf {\bibinfo {volume} {105}},\ \bibinfo {pages} {094028} (\bibinfo {year} {2022})},\ \Eprint {http://arxiv.org/abs/2201.06428} {arXiv:2201.06428 [hep-ph]} \BibitemShut {NoStop}%
\bibitem [{\citenamefont {Lan}\ \emph {et~al.}(2022)\citenamefont {Lan}, \citenamefont {Fu}, \citenamefont {Mondal}, \citenamefont {Zhao},\ and\ \citenamefont {Vary}}]{Lan:2021wok}%
  \BibitemOpen
  \bibfield  {author} {\bibinfo {author} {\bibfnamefont {J.}~\bibnamefont {Lan}}, \bibinfo {author} {\bibfnamefont {K.}~\bibnamefont {Fu}}, \bibinfo {author} {\bibfnamefont {C.}~\bibnamefont {Mondal}}, \bibinfo {author} {\bibfnamefont {X.}~\bibnamefont {Zhao}}, \ and\ \bibinfo {author} {\bibfnamefont {j.~P.}\ \bibnamefont {Vary}} (\bibinfo {collaboration} {BLFQ}),\ }\href {\doibase 10.1016/j.physletb.2022.136890} {\bibfield  {journal} {\bibinfo  {journal} {Phys. Lett. B}\ }\textbf {\bibinfo {volume} {825}},\ \bibinfo {pages} {136890} (\bibinfo {year} {2022})},\ \Eprint {http://arxiv.org/abs/2106.04954} {arXiv:2106.04954 [hep-ph]} \BibitemShut {NoStop}%
\bibitem [{\citenamefont {Xu}\ \emph {et~al.}(2023)\citenamefont {Xu}, \citenamefont {Mondal}, \citenamefont {Zhao}, \citenamefont {Li},\ and\ \citenamefont {Vary}}]{Xu:2023nqv}%
  \BibitemOpen
  \bibfield  {author} {\bibinfo {author} {\bibfnamefont {S.}~\bibnamefont {Xu}}, \bibinfo {author} {\bibfnamefont {C.}~\bibnamefont {Mondal}}, \bibinfo {author} {\bibfnamefont {X.}~\bibnamefont {Zhao}}, \bibinfo {author} {\bibfnamefont {Y.}~\bibnamefont {Li}}, \ and\ \bibinfo {author} {\bibfnamefont {J.~P.}\ \bibnamefont {Vary}} (\bibinfo {collaboration} {BLFQ}),\ }\href {\doibase 10.1103/PhysRevD.108.094002} {\bibfield  {journal} {\bibinfo  {journal} {Phys. Rev. D}\ }\textbf {\bibinfo {volume} {108}},\ \bibinfo {pages} {094002} (\bibinfo {year} {2023})}\BibitemShut {NoStop}%
\bibitem [{\citenamefont {Adamczyk}\ \emph {et~al.}(2015)\citenamefont {Adamczyk} \emph {et~al.}}]{STAR:2014wox}%
  \BibitemOpen
  \bibfield  {author} {\bibinfo {author} {\bibfnamefont {L.}~\bibnamefont {Adamczyk}} \emph {et~al.} (\bibinfo {collaboration} {STAR}),\ }\href {\doibase 10.1103/PhysRevLett.115.092002} {\bibfield  {journal} {\bibinfo  {journal} {Phys. Rev. Lett.}\ }\textbf {\bibinfo {volume} {115}},\ \bibinfo {pages} {092002} (\bibinfo {year} {2015})},\ \Eprint {http://arxiv.org/abs/1405.5134} {arXiv:1405.5134 [hep-ex]} \BibitemShut {NoStop}%
\bibitem [{\citenamefont {de~Florian}\ \emph {et~al.}(2014)\citenamefont {de~Florian}, \citenamefont {Sassot}, \citenamefont {Stratmann},\ and\ \citenamefont {Vogelsang}}]{deFlorian:2014yva}%
  \BibitemOpen
  \bibfield  {author} {\bibinfo {author} {\bibfnamefont {D.}~\bibnamefont {de~Florian}}, \bibinfo {author} {\bibfnamefont {R.}~\bibnamefont {Sassot}}, \bibinfo {author} {\bibfnamefont {M.}~\bibnamefont {Stratmann}}, \ and\ \bibinfo {author} {\bibfnamefont {W.}~\bibnamefont {Vogelsang}},\ }\href {\doibase 10.1103/PhysRevLett.113.012001} {\bibfield  {journal} {\bibinfo  {journal} {Phys. Rev. Lett.}\ }\textbf {\bibinfo {volume} {113}},\ \bibinfo {pages} {012001} (\bibinfo {year} {2014})},\ \Eprint {http://arxiv.org/abs/1404.4293} {arXiv:1404.4293 [hep-ph]} \BibitemShut {NoStop}%
\bibitem [{\citenamefont {Nocera}\ \emph {et~al.}(2014)\citenamefont {Nocera}, \citenamefont {Ball}, \citenamefont {Forte}, \citenamefont {Ridolfi},\ and\ \citenamefont {Rojo}}]{Nocera:2014gqa}%
  \BibitemOpen
  \bibfield  {author} {\bibinfo {author} {\bibfnamefont {E.~R.}\ \bibnamefont {Nocera}}, \bibinfo {author} {\bibfnamefont {R.~D.}\ \bibnamefont {Ball}}, \bibinfo {author} {\bibfnamefont {S.}~\bibnamefont {Forte}}, \bibinfo {author} {\bibfnamefont {G.}~\bibnamefont {Ridolfi}}, \ and\ \bibinfo {author} {\bibfnamefont {J.}~\bibnamefont {Rojo}} (\bibinfo {collaboration} {NNPDF}),\ }\href {\doibase 10.1016/j.nuclphysb.2014.08.008} {\bibfield  {journal} {\bibinfo  {journal} {Nuclear Physics B}\ }\textbf {\bibinfo {volume} {887}},\ \bibinfo {pages} {276} (\bibinfo {year} {2014})},\ \Eprint {http://arxiv.org/abs/1406.5539} {arXiv:1406.5539 [hep-ph]} \BibitemShut {NoStop}%
\bibitem [{\citenamefont {Ethier}\ \emph {et~al.}(2017)\citenamefont {Ethier}, \citenamefont {Sato},\ and\ \citenamefont {Melnitchouk}}]{Ethier:2017zbq}%
  \BibitemOpen
  \bibfield  {author} {\bibinfo {author} {\bibfnamefont {J.~J.}\ \bibnamefont {Ethier}}, \bibinfo {author} {\bibfnamefont {N.}~\bibnamefont {Sato}}, \ and\ \bibinfo {author} {\bibfnamefont {W.}~\bibnamefont {Melnitchouk}},\ }\href {\doibase 10.1103/PhysRevLett.119.132001} {\bibfield  {journal} {\bibinfo  {journal} {Phys. Rev. Lett.}\ }\textbf {\bibinfo {volume} {119}},\ \bibinfo {pages} {132001} (\bibinfo {year} {2017})},\ \Eprint {http://arxiv.org/abs/1705.05889} {arXiv:1705.05889 [hep-ph]} \BibitemShut {NoStop}%
\bibitem [{\citenamefont {Ji}\ \emph {et~al.}(2021)\citenamefont {Ji}, \citenamefont {Yuan},\ and\ \citenamefont {Zhao}}]{Ji:2020ena}%
  \BibitemOpen
  \bibfield  {author} {\bibinfo {author} {\bibfnamefont {X.}~\bibnamefont {Ji}}, \bibinfo {author} {\bibfnamefont {F.}~\bibnamefont {Yuan}}, \ and\ \bibinfo {author} {\bibfnamefont {Y.}~\bibnamefont {Zhao}},\ }\href {\doibase 10.1038/s42254-020-00248-4} {\bibfield  {journal} {\bibinfo  {journal} {Nature Rev. Phys.}\ }\textbf {\bibinfo {volume} {3}},\ \bibinfo {pages} {27} (\bibinfo {year} {2021})},\ \Eprint {http://arxiv.org/abs/2009.01291} {arXiv:2009.01291 [hep-ph]} \BibitemShut {NoStop}%
\bibitem [{\citenamefont {Accardi}\ \emph {et~al.}(2016)\citenamefont {Accardi} \emph {et~al.}}]{Accardi:2012qut}%
  \BibitemOpen
  \bibfield  {author} {\bibinfo {author} {\bibfnamefont {A.}~\bibnamefont {Accardi}} \emph {et~al.},\ }\href {\doibase 10.1140/epja/i2016-16268-9} {\bibfield  {journal} {\bibinfo  {journal} {Eur. Phys. J. A}\ }\textbf {\bibinfo {volume} {52}},\ \bibinfo {pages} {268} (\bibinfo {year} {2016})},\ \Eprint {http://arxiv.org/abs/1212.1701} {arXiv:1212.1701 [nucl-ex]} \BibitemShut {NoStop}%
\bibitem [{\citenamefont {Abdul~Khalek}\ \emph {et~al.}(2022)\citenamefont {Abdul~Khalek} \emph {et~al.}}]{AbdulKhalek:2021gbh}%
  \BibitemOpen
  \bibfield  {author} {\bibinfo {author} {\bibfnamefont {R.}~\bibnamefont {Abdul~Khalek}} \emph {et~al.},\ }\href {\doibase 10.1016/j.nuclphysa.2022.122447} {\bibfield  {journal} {\bibinfo  {journal} {Nucl. Phys. A}\ }\textbf {\bibinfo {volume} {1026}},\ \bibinfo {pages} {122447} (\bibinfo {year} {2022})},\ \Eprint {http://arxiv.org/abs/2103.05419} {arXiv:2103.05419 [physics.ins-det]} \BibitemShut {NoStop}%
\bibitem [{\citenamefont {Anderle}\ \emph {et~al.}(2021)\citenamefont {Anderle} \emph {et~al.}}]{Anderle:2021wcy}%
  \BibitemOpen
  \bibfield  {author} {\bibinfo {author} {\bibfnamefont {D.~P.}\ \bibnamefont {Anderle}} \emph {et~al.},\ }\href {\doibase 10.1007/s11467-021-1062-0} {\bibfield  {journal} {\bibinfo  {journal} {Front. Phys. (Beijing)}\ }\textbf {\bibinfo {volume} {16}},\ \bibinfo {pages} {64701} (\bibinfo {year} {2021})},\ \Eprint {http://arxiv.org/abs/2102.09222} {arXiv:2102.09222 [nucl-ex]} \BibitemShut {NoStop}%
\bibitem [{\citenamefont {Zhao}(2015)}]{Zhao:2014hpa}%
  \BibitemOpen
  \bibfield  {author} {\bibinfo {author} {\bibfnamefont {X.}~\bibnamefont {Zhao}},\ }\href {\doibase 10.1007/s00601-015-1003-y} {\bibfield  {journal} {\bibinfo  {journal} {Few Body Syst.}\ }\textbf {\bibinfo {volume} {56}},\ \bibinfo {pages} {257} (\bibinfo {year} {2015})},\ \Eprint {http://arxiv.org/abs/1411.7748} {arXiv:1411.7748 [nucl-th]} \BibitemShut {NoStop}%
\bibitem [{\citenamefont {Zhao}\ \emph {et~al.}(2020)\citenamefont {Zhao}, \citenamefont {Fu}, \citenamefont {Zhao},\ and\ \citenamefont {Vary}}]{Zhao:2020kuf}%
  \BibitemOpen
  \bibfield  {author} {\bibinfo {author} {\bibfnamefont {X.}~\bibnamefont {Zhao}}, \bibinfo {author} {\bibfnamefont {K.}~\bibnamefont {Fu}}, \bibinfo {author} {\bibfnamefont {H.}~\bibnamefont {Zhao}}, \ and\ \bibinfo {author} {\bibfnamefont {J.~P.}\ \bibnamefont {Vary}},\ }\href {\doibase 10.22323/1.374.0090} {\bibfield  {journal} {\bibinfo  {journal} {PoS}\ }\textbf {\bibinfo {volume} {LC2019}},\ \bibinfo {pages} {090} (\bibinfo {year} {2020})},\ \Eprint {http://arxiv.org/abs/2103.06719} {arXiv:2103.06719 [hep-ph]} \BibitemShut {NoStop}%
\bibitem [{\citenamefont {Burkardt}(1998)}]{Burkardt:1998dd}%
  \BibitemOpen
  \bibfield  {author} {\bibinfo {author} {\bibfnamefont {M.}~\bibnamefont {Burkardt}},\ }\href {\doibase 10.1103/PhysRevD.58.096015} {\bibfield  {journal} {\bibinfo  {journal} {Phys. Rev. D}\ }\textbf {\bibinfo {volume} {58}},\ \bibinfo {pages} {096015} (\bibinfo {year} {1998})},\ \Eprint {http://arxiv.org/abs/hep-th/9805088} {arXiv:hep-th/9805088} \BibitemShut {NoStop}%
\bibitem [{\citenamefont {Glazek}\ and\ \citenamefont {Perry}(1992)}]{Glazek:1992aq}%
  \BibitemOpen
  \bibfield  {author} {\bibinfo {author} {\bibfnamefont {S.~D.}\ \bibnamefont {Glazek}}\ and\ \bibinfo {author} {\bibfnamefont {R.~J.}\ \bibnamefont {Perry}},\ }\href {\doibase 10.1103/PhysRevD.45.3740} {\bibfield  {journal} {\bibinfo  {journal} {Phys. Rev. D}\ }\textbf {\bibinfo {volume} {45}},\ \bibinfo {pages} {3740} (\bibinfo {year} {1992})}\BibitemShut {NoStop}%
\bibitem [{\citenamefont {Brodsky}\ \emph {et~al.}(2001)\citenamefont {Brodsky}, \citenamefont {Diehl},\ and\ \citenamefont {Hwang}}]{Brodsky:2000xy}%
  \BibitemOpen
  \bibfield  {author} {\bibinfo {author} {\bibfnamefont {S.~J.}\ \bibnamefont {Brodsky}}, \bibinfo {author} {\bibfnamefont {M.}~\bibnamefont {Diehl}}, \ and\ \bibinfo {author} {\bibfnamefont {D.~S.}\ \bibnamefont {Hwang}},\ }\href {\doibase 10.1016/S0550-3213(00)00695-7} {\bibfield  {journal} {\bibinfo  {journal} {Nucl. Phys. B}\ }\textbf {\bibinfo {volume} {596}},\ \bibinfo {pages} {99} (\bibinfo {year} {2001})},\ \Eprint {http://arxiv.org/abs/hep-ph/0009254} {arXiv:hep-ph/0009254} \BibitemShut {NoStop}%
\bibitem [{\citenamefont {Cates}\ \emph {et~al.}(2011)\citenamefont {Cates}, \citenamefont {de~Jager}, \citenamefont {Riordan},\ and\ \citenamefont {Wojtsekhowski}}]{Cates:2011pz}%
  \BibitemOpen
  \bibfield  {author} {\bibinfo {author} {\bibfnamefont {G.~D.}\ \bibnamefont {Cates}}, \bibinfo {author} {\bibfnamefont {C.~W.}\ \bibnamefont {de~Jager}}, \bibinfo {author} {\bibfnamefont {S.}~\bibnamefont {Riordan}}, \ and\ \bibinfo {author} {\bibfnamefont {B.}~\bibnamefont {Wojtsekhowski}},\ }\href {\doibase 10.1103/PhysRevLett.106.252003} {\bibfield  {journal} {\bibinfo  {journal} {Phys. Rev. Lett.}\ }\textbf {\bibinfo {volume} {106}},\ \bibinfo {pages} {252003} (\bibinfo {year} {2011})},\ \Eprint {http://arxiv.org/abs/1103.1808} {arXiv:1103.1808 [nucl-ex]} \BibitemShut {NoStop}%
\bibitem [{\citenamefont {Gayou}\ \emph {et~al.}(2001)\citenamefont {Gayou} \emph {et~al.}}]{Gayou:2001qt}%
  \BibitemOpen
  \bibfield  {author} {\bibinfo {author} {\bibfnamefont {O.}~\bibnamefont {Gayou}} \emph {et~al.},\ }\href {\doibase 10.1103/PhysRevC.64.038202} {\bibfield  {journal} {\bibinfo  {journal} {Phys. Rev. C}\ }\textbf {\bibinfo {volume} {64}},\ \bibinfo {pages} {038202} (\bibinfo {year} {2001})}\BibitemShut {NoStop}%
\bibitem [{\citenamefont {Jones}\ \emph {et~al.}(2000)\citenamefont {Jones} \emph {et~al.}}]{JeffersonLabHallA:1999epl}%
  \BibitemOpen
  \bibfield  {author} {\bibinfo {author} {\bibfnamefont {M.~K.}\ \bibnamefont {Jones}} \emph {et~al.} (\bibinfo {collaboration} {Jefferson Lab Hall A}),\ }\href {\doibase 10.1103/PhysRevLett.84.1398} {\bibfield  {journal} {\bibinfo  {journal} {Phys. Rev. Lett.}\ }\textbf {\bibinfo {volume} {84}},\ \bibinfo {pages} {1398} (\bibinfo {year} {2000})},\ \Eprint {http://arxiv.org/abs/nucl-ex/9910005} {arXiv:nucl-ex/9910005} \BibitemShut {NoStop}%
\bibitem [{\citenamefont {Arrington}\ \emph {et~al.}(2007)\citenamefont {Arrington}, \citenamefont {Melnitchouk},\ and\ \citenamefont {Tjon}}]{Arrington:2007ux}%
  \BibitemOpen
  \bibfield  {author} {\bibinfo {author} {\bibfnamefont {J.}~\bibnamefont {Arrington}}, \bibinfo {author} {\bibfnamefont {W.}~\bibnamefont {Melnitchouk}}, \ and\ \bibinfo {author} {\bibfnamefont {J.~A.}\ \bibnamefont {Tjon}},\ }\href {\doibase 10.1103/PhysRevC.76.035205} {\bibfield  {journal} {\bibinfo  {journal} {Phys. Rev. C}\ }\textbf {\bibinfo {volume} {76}},\ \bibinfo {pages} {035205} (\bibinfo {year} {2007})},\ \Eprint {http://arxiv.org/abs/0707.1861} {arXiv:0707.1861 [nucl-ex]} \BibitemShut {NoStop}%
\bibitem [{\citenamefont {Gayou}\ \emph {et~al.}(2002)\citenamefont {Gayou} \emph {et~al.}}]{JeffersonLabHallA:2001qqe}%
  \BibitemOpen
  \bibfield  {author} {\bibinfo {author} {\bibfnamefont {O.}~\bibnamefont {Gayou}} \emph {et~al.} (\bibinfo {collaboration} {Jefferson Lab Hall A}),\ }\href {\doibase 10.1103/PhysRevLett.88.092301} {\bibfield  {journal} {\bibinfo  {journal} {Phys. Rev. Lett.}\ }\textbf {\bibinfo {volume} {88}},\ \bibinfo {pages} {092301} (\bibinfo {year} {2002})},\ \Eprint {http://arxiv.org/abs/nucl-ex/0111010} {arXiv:nucl-ex/0111010} \BibitemShut {NoStop}%
\bibitem [{\citenamefont {Pospischil}\ \emph {et~al.}(2001)\citenamefont {Pospischil} \emph {et~al.}}]{A1:2001xxy}%
  \BibitemOpen
  \bibfield  {author} {\bibinfo {author} {\bibfnamefont {T.}~\bibnamefont {Pospischil}} \emph {et~al.} (\bibinfo {collaboration} {A1}),\ }\href {\doibase 10.1007/s100500170046} {\bibfield  {journal} {\bibinfo  {journal} {Eur. Phys. J. A}\ }\textbf {\bibinfo {volume} {12}},\ \bibinfo {pages} {125} (\bibinfo {year} {2001})}\BibitemShut {NoStop}%
\bibitem [{\citenamefont {Milbrath}\ \emph {et~al.}(1998)\citenamefont {Milbrath} \emph {et~al.}}]{BatesFPP:1997rpw}%
  \BibitemOpen
  \bibfield  {author} {\bibinfo {author} {\bibfnamefont {B.~D.}\ \bibnamefont {Milbrath}} \emph {et~al.} (\bibinfo {collaboration} {Bates FPP}),\ }\href {\doibase 10.1103/PhysRevLett.80.452} {\bibfield  {journal} {\bibinfo  {journal} {Phys. Rev. Lett.}\ }\textbf {\bibinfo {volume} {80}},\ \bibinfo {pages} {452} (\bibinfo {year} {1998})},\ \bibinfo {note} {[Erratum: Phys.Rev.Lett. 82, 2221 (1999)]},\ \Eprint {http://arxiv.org/abs/nucl-ex/9712006} {arXiv:nucl-ex/9712006} \BibitemShut {NoStop}%
\bibitem [{\citenamefont {Arrington}(2005)}]{Arrington:2004ae}%
  \BibitemOpen
  \bibfield  {author} {\bibinfo {author} {\bibfnamefont {J.}~\bibnamefont {Arrington}},\ }\href {\doibase 10.1103/PhysRevC.71.015202} {\bibfield  {journal} {\bibinfo  {journal} {Phys. Rev. C}\ }\textbf {\bibinfo {volume} {71}},\ \bibinfo {pages} {015202} (\bibinfo {year} {2005})},\ \Eprint {http://arxiv.org/abs/hep-ph/0408261} {arXiv:hep-ph/0408261} \BibitemShut {NoStop}%
\bibitem [{\citenamefont {Ernst}\ \emph {et~al.}(1960)\citenamefont {Ernst}, \citenamefont {Sachs},\ and\ \citenamefont {Wali}}]{Ernst:1960zza}%
  \BibitemOpen
  \bibfield  {author} {\bibinfo {author} {\bibfnamefont {F.~J.}\ \bibnamefont {Ernst}}, \bibinfo {author} {\bibfnamefont {R.~G.}\ \bibnamefont {Sachs}}, \ and\ \bibinfo {author} {\bibfnamefont {K.~C.}\ \bibnamefont {Wali}},\ }\href {\doibase 10.1103/PhysRev.119.1105} {\bibfield  {journal} {\bibinfo  {journal} {Phys. Rev.}\ }\textbf {\bibinfo {volume} {119}},\ \bibinfo {pages} {1105} (\bibinfo {year} {1960})}\BibitemShut {NoStop}%
\bibitem [{\citenamefont {Lin}\ \emph {et~al.}(2022)\citenamefont {Lin}, \citenamefont {Hammer},\ and\ \citenamefont {Mei\ss{}ner}}]{Lin:2021xrc}%
  \BibitemOpen
  \bibfield  {author} {\bibinfo {author} {\bibfnamefont {Y.-H.}\ \bibnamefont {Lin}}, \bibinfo {author} {\bibfnamefont {H.-W.}\ \bibnamefont {Hammer}}, \ and\ \bibinfo {author} {\bibfnamefont {U.-G.}\ \bibnamefont {Mei\ss{}ner}},\ }\href {\doibase 10.1103/PhysRevLett.128.052002} {\bibfield  {journal} {\bibinfo  {journal} {Phys. Rev. Lett.}\ }\textbf {\bibinfo {volume} {128}},\ \bibinfo {pages} {052002} (\bibinfo {year} {2022})},\ \Eprint {http://arxiv.org/abs/2109.12961} {arXiv:2109.12961 [hep-ph]} \BibitemShut {NoStop}%
\bibitem [{\citenamefont {Navas}\ \emph {et~al.}(2024)\citenamefont {Navas} \emph {et~al.}}]{ParticleDataGroup:2024cfk}%
  \BibitemOpen
  \bibfield  {author} {\bibinfo {author} {\bibfnamefont {S.}~\bibnamefont {Navas}} \emph {et~al.} (\bibinfo {collaboration} {Particle Data Group}),\ }\href {\doibase 10.1103/PhysRevD.110.030001} {\bibfield  {journal} {\bibinfo  {journal} {Phys. Rev. D}\ }\textbf {\bibinfo {volume} {110}},\ \bibinfo {pages} {030001} (\bibinfo {year} {2024})}\BibitemShut {NoStop}%
\bibitem [{\citenamefont {Ball}\ \emph {et~al.}(2017)\citenamefont {Ball}, \citenamefont {Bertone}, \citenamefont {Carrazza}, \citenamefont {Debbio}, \citenamefont {Forte}, \citenamefont {{Groth-Merrild}}, \citenamefont {Guffanti}, \citenamefont {Hartland}, \citenamefont {Kassabov}, \citenamefont {Latorre}, \citenamefont {Nocera}, \citenamefont {Rojo}, \citenamefont {Rottoli}, \citenamefont {Slade},\ and\ \citenamefont {Ubiali}}]{NNPDF:2017mvq}%
  \BibitemOpen
  \bibfield  {author} {\bibinfo {author} {\bibfnamefont {R.~D.}\ \bibnamefont {Ball}}, \bibinfo {author} {\bibfnamefont {V.}~\bibnamefont {Bertone}}, \bibinfo {author} {\bibfnamefont {S.}~\bibnamefont {Carrazza}}, \bibinfo {author} {\bibfnamefont {L.~D.}\ \bibnamefont {Debbio}}, \bibinfo {author} {\bibfnamefont {S.}~\bibnamefont {Forte}}, \bibinfo {author} {\bibfnamefont {P.}~\bibnamefont {{Groth-Merrild}}}, \bibinfo {author} {\bibfnamefont {A.}~\bibnamefont {Guffanti}}, \bibinfo {author} {\bibfnamefont {N.~P.}\ \bibnamefont {Hartland}}, \bibinfo {author} {\bibfnamefont {Z.}~\bibnamefont {Kassabov}}, \bibinfo {author} {\bibfnamefont {J.~I.}\ \bibnamefont {Latorre}}, \bibinfo {author} {\bibfnamefont {E.~R.}\ \bibnamefont {Nocera}}, \bibinfo {author} {\bibfnamefont {J.}~\bibnamefont {Rojo}}, \bibinfo {author} {\bibfnamefont {L.}~\bibnamefont {Rottoli}}, \bibinfo {author} {\bibfnamefont {E.}~\bibnamefont {Slade}}, \ and\ \bibinfo {author} {\bibfnamefont {M.}~\bibnamefont {Ubiali}} (\bibinfo {collaboration}
  {NNPDF}),\ }\href {\doibase 10.1140/epjc/s10052-017-5199-5} {\bibfield  {journal} {\bibinfo  {journal} {The European Physical Journal C}\ }\textbf {\bibinfo {volume} {77}},\ \bibinfo {pages} {663} (\bibinfo {year} {2017})},\ \Eprint {http://arxiv.org/abs/1706.00428} {arXiv:1706.00428 [hep-ph]} \BibitemShut {NoStop}%
\bibitem [{\citenamefont {Cocuzza}\ \emph {et~al.}(2021)\citenamefont {Cocuzza}, \citenamefont {Melnitchouk}, \citenamefont {Metz},\ and\ \citenamefont {Sato}}]{Cocuzza:2021cbi}%
  \BibitemOpen
  \bibfield  {author} {\bibinfo {author} {\bibfnamefont {C.}~\bibnamefont {Cocuzza}}, \bibinfo {author} {\bibfnamefont {W.}~\bibnamefont {Melnitchouk}}, \bibinfo {author} {\bibfnamefont {A.}~\bibnamefont {Metz}}, \ and\ \bibinfo {author} {\bibfnamefont {N.}~\bibnamefont {Sato}} (\bibinfo {collaboration} {Jefferson Lab Angular Momentum (JAM)}),\ }\href {\doibase 10.1103/PhysRevD.104.074031} {\bibfield  {journal} {\bibinfo  {journal} {Physical Review D}\ }\textbf {\bibinfo {volume} {104}},\ \bibinfo {pages} {074031} (\bibinfo {year} {2021})},\ \Eprint {http://arxiv.org/abs/2109.00677} {arXiv:2109.00677 [hep-ph]} \BibitemShut {NoStop}%
\bibitem [{\citenamefont {Salam}\ and\ \citenamefont {Rojo}(2009)}]{Salam:2008qg}%
  \BibitemOpen
  \bibfield  {author} {\bibinfo {author} {\bibfnamefont {G.~P.}\ \bibnamefont {Salam}}\ and\ \bibinfo {author} {\bibfnamefont {J.}~\bibnamefont {Rojo}},\ }\href {\doibase 10.1016/j.cpc.2008.08.010} {\bibfield  {journal} {\bibinfo  {journal} {Comput. Phys. Commun.}\ }\textbf {\bibinfo {volume} {180}},\ \bibinfo {pages} {120} (\bibinfo {year} {2009})},\ \Eprint {http://arxiv.org/abs/0804.3755} {arXiv:0804.3755 [hep-ph]} \BibitemShut {NoStop}%
\bibitem [{\citenamefont {Dokshitzer}(1977)}]{Dokshitzer:1977sg}%
  \BibitemOpen
  \bibfield  {author} {\bibinfo {author} {\bibfnamefont {Y.~L.}\ \bibnamefont {Dokshitzer}},\ }\href@noop {} {\bibfield  {journal} {\bibinfo  {journal} {Sov. Phys. JETP}\ }\textbf {\bibinfo {volume} {46}},\ \bibinfo {pages} {641} (\bibinfo {year} {1977})}\BibitemShut {NoStop}%
\bibitem [{\citenamefont {Gribov}\ and\ \citenamefont {Lipatov}(1972)}]{Gribov:1972ri}%
  \BibitemOpen
  \bibfield  {author} {\bibinfo {author} {\bibfnamefont {V.~N.}\ \bibnamefont {Gribov}}\ and\ \bibinfo {author} {\bibfnamefont {L.~N.}\ \bibnamefont {Lipatov}},\ }\href@noop {} {\bibfield  {journal} {\bibinfo  {journal} {Sov. J. Nucl. Phys.}\ }\textbf {\bibinfo {volume} {15}},\ \bibinfo {pages} {438} (\bibinfo {year} {1972})}\BibitemShut {NoStop}%
\bibitem [{\citenamefont {Altarelli}\ and\ \citenamefont {Parisi}(1977)}]{Altarelli:1977zs}%
  \BibitemOpen
  \bibfield  {author} {\bibinfo {author} {\bibfnamefont {G.}~\bibnamefont {Altarelli}}\ and\ \bibinfo {author} {\bibfnamefont {G.}~\bibnamefont {Parisi}},\ }\href {\doibase 10.1016/0550-3213(77)90384-4} {\bibfield  {journal} {\bibinfo  {journal} {Nucl. Phys. B}\ }\textbf {\bibinfo {volume} {126}},\ \bibinfo {pages} {298} (\bibinfo {year} {1977})}\BibitemShut {NoStop}%
\bibitem [{\citenamefont {Alekseev}\ \emph {et~al.}(2010)\citenamefont {Alekseev} \emph {et~al.}}]{COMPASS:2010hwr}%
  \BibitemOpen
  \bibfield  {author} {\bibinfo {author} {\bibfnamefont {M.~G.}\ \bibnamefont {Alekseev}} \emph {et~al.} (\bibinfo {collaboration} {COMPASS}),\ }\href {\doibase 10.1016/j.physletb.2010.08.034} {\bibfield  {journal} {\bibinfo  {journal} {Phys. Lett. B}\ }\textbf {\bibinfo {volume} {693}},\ \bibinfo {pages} {227} (\bibinfo {year} {2010})},\ \Eprint {http://arxiv.org/abs/1007.4061} {arXiv:1007.4061 [hep-ex]} \BibitemShut {NoStop}%
\bibitem [{\citenamefont {Airapetian}\ \emph {et~al.}(2004)\citenamefont {Airapetian} \emph {et~al.}}]{HERMES:2003gbu}%
  \BibitemOpen
  \bibfield  {author} {\bibinfo {author} {\bibfnamefont {A.}~\bibnamefont {Airapetian}} \emph {et~al.} (\bibinfo {collaboration} {HERMES}),\ }\href {\doibase 10.1103/PhysRevLett.92.012005} {\bibfield  {journal} {\bibinfo  {journal} {Phys. Rev. Lett.}\ }\textbf {\bibinfo {volume} {92}},\ \bibinfo {pages} {012005} (\bibinfo {year} {2004})},\ \Eprint {http://arxiv.org/abs/hep-ex/0307064} {arXiv:hep-ex/0307064} \BibitemShut {NoStop}%
\bibitem [{\citenamefont {Airapetian}\ \emph {et~al.}(2005)\citenamefont {Airapetian} \emph {et~al.}}]{HERMES:2004zsh}%
  \BibitemOpen
  \bibfield  {author} {\bibinfo {author} {\bibfnamefont {A.}~\bibnamefont {Airapetian}} \emph {et~al.} (\bibinfo {collaboration} {HERMES}),\ }\href {\doibase 10.1103/PhysRevD.71.012003} {\bibfield  {journal} {\bibinfo  {journal} {Phys. Rev. D}\ }\textbf {\bibinfo {volume} {71}},\ \bibinfo {pages} {012003} (\bibinfo {year} {2005})},\ \Eprint {http://arxiv.org/abs/hep-ex/0407032} {arXiv:hep-ex/0407032} \BibitemShut {NoStop}%
\bibitem [{\citenamefont {Sato}\ \emph {et~al.}(2016)\citenamefont {Sato}, \citenamefont {Melnitchouk}, \citenamefont {Kuhn}, \citenamefont {Ethier},\ and\ \citenamefont {Accardi}}]{Sato:2016tuz}%
  \BibitemOpen
  \bibfield  {author} {\bibinfo {author} {\bibfnamefont {N.}~\bibnamefont {Sato}}, \bibinfo {author} {\bibfnamefont {W.}~\bibnamefont {Melnitchouk}}, \bibinfo {author} {\bibfnamefont {S.~E.}\ \bibnamefont {Kuhn}}, \bibinfo {author} {\bibfnamefont {J.~J.}\ \bibnamefont {Ethier}}, \ and\ \bibinfo {author} {\bibfnamefont {A.}~\bibnamefont {Accardi}} (\bibinfo {collaboration} {Jefferson Lab Angular Momentum}),\ }\href {\doibase 10.1103/PhysRevD.93.074005} {\bibfield  {journal} {\bibinfo  {journal} {Physical Review D}\ }\textbf {\bibinfo {volume} {93}},\ \bibinfo {pages} {074005} (\bibinfo {year} {2016})},\ \Eprint {http://arxiv.org/abs/1601.07782} {arXiv:1601.07782 [hep-ph]} \BibitemShut {NoStop}%
\bibitem [{\citenamefont {Zhou}\ \emph {et~al.}(2022)\citenamefont {Zhou}, \citenamefont {Sato},\ and\ \citenamefont {Melnitchouk}}]{Zhou:2022wzm}%
  \BibitemOpen
  \bibfield  {author} {\bibinfo {author} {\bibfnamefont {Y.}~\bibnamefont {Zhou}}, \bibinfo {author} {\bibfnamefont {N.}~\bibnamefont {Sato}}, \ and\ \bibinfo {author} {\bibfnamefont {W.}~\bibnamefont {Melnitchouk}} (\bibinfo {collaboration} {Jefferson Lab Angular Momentum (JAM)}),\ }\href {\doibase 10.1103/PhysRevD.105.074022} {\bibfield  {journal} {\bibinfo  {journal} {Phys. Rev. D}\ }\textbf {\bibinfo {volume} {105}},\ \bibinfo {pages} {074022} (\bibinfo {year} {2022})},\ \Eprint {http://arxiv.org/abs/2201.02075} {arXiv:2201.02075 [hep-ph]} \BibitemShut {NoStop}%
\bibitem [{\citenamefont {Deur}\ \emph {et~al.}(2019)\citenamefont {Deur}, \citenamefont {Brodsky},\ and\ \citenamefont {De~T{\'e}ramond}}]{Deur:2018roz}%
  \BibitemOpen
  \bibfield  {author} {\bibinfo {author} {\bibfnamefont {A.}~\bibnamefont {Deur}}, \bibinfo {author} {\bibfnamefont {S.~J.}\ \bibnamefont {Brodsky}}, \ and\ \bibinfo {author} {\bibfnamefont {G.~F.}\ \bibnamefont {De~T{\'e}ramond}},\ }\href {\doibase 10.1088/1361-6633/ab0b8f} {\bibfield  {journal} {\bibinfo  {journal} {Reports on Progress in Physics}\ }\textbf {\bibinfo {volume} {82}},\ \bibinfo {pages} {076201} (\bibinfo {year} {2019})},\ \Eprint {http://arxiv.org/abs/1807.05250} {arXiv:1807.05250 [hep-ph]} \BibitemShut {NoStop}%
\bibitem [{\citenamefont {Yang}\ \emph {et~al.}(2017)\citenamefont {Yang}, \citenamefont {Sufian}, \citenamefont {Alexandru}, \citenamefont {Draper}, \citenamefont {Glatzmaier}, \citenamefont {Liu},\ and\ \citenamefont {Zhao}}]{Yang:2016plb}%
  \BibitemOpen
  \bibfield  {author} {\bibinfo {author} {\bibfnamefont {Y.-B.}\ \bibnamefont {Yang}}, \bibinfo {author} {\bibfnamefont {R.~S.}\ \bibnamefont {Sufian}}, \bibinfo {author} {\bibfnamefont {A.}~\bibnamefont {Alexandru}}, \bibinfo {author} {\bibfnamefont {T.}~\bibnamefont {Draper}}, \bibinfo {author} {\bibfnamefont {M.~J.}\ \bibnamefont {Glatzmaier}}, \bibinfo {author} {\bibfnamefont {K.-F.}\ \bibnamefont {Liu}}, \ and\ \bibinfo {author} {\bibfnamefont {Y.}~\bibnamefont {Zhao}},\ }\href {\doibase 10.1103/PhysRevLett.118.102001} {\bibfield  {journal} {\bibinfo  {journal} {Phys. Rev. Lett.}\ }\textbf {\bibinfo {volume} {118}},\ \bibinfo {pages} {102001} (\bibinfo {year} {2017})},\ \Eprint {http://arxiv.org/abs/1609.05937} {arXiv:1609.05937 [hep-ph]} \BibitemShut {NoStop}%
\bibitem [{\citenamefont {Ji}(1997)}]{Ji:1996ek}%
  \BibitemOpen
  \bibfield  {author} {\bibinfo {author} {\bibfnamefont {X.-D.}\ \bibnamefont {Ji}},\ }\href {\doibase 10.1103/PhysRevLett.78.610} {\bibfield  {journal} {\bibinfo  {journal} {Phys. Rev. Lett.}\ }\textbf {\bibinfo {volume} {78}},\ \bibinfo {pages} {610} (\bibinfo {year} {1997})},\ \Eprint {http://arxiv.org/abs/hep-ph/9603249} {arXiv:hep-ph/9603249} \BibitemShut {NoStop}%
\bibitem [{\citenamefont {Bhattacharya}\ \emph {et~al.}(2022)\citenamefont {Bhattacharya}, \citenamefont {Boussarie},\ and\ \citenamefont {Hatta}}]{Bhattacharya:2022vvo}%
  \BibitemOpen
  \bibfield  {author} {\bibinfo {author} {\bibfnamefont {S.}~\bibnamefont {Bhattacharya}}, \bibinfo {author} {\bibfnamefont {R.}~\bibnamefont {Boussarie}}, \ and\ \bibinfo {author} {\bibfnamefont {Y.}~\bibnamefont {Hatta}},\ }\href {\doibase 10.1103/PhysRevLett.128.182002} {\bibfield  {journal} {\bibinfo  {journal} {Phys. Rev. Lett.}\ }\textbf {\bibinfo {volume} {128}},\ \bibinfo {pages} {182002} (\bibinfo {year} {2022})},\ \Eprint {http://arxiv.org/abs/2201.08709} {arXiv:2201.08709 [hep-ph]} \BibitemShut {NoStop}%
\bibitem [{\citenamefont {Bhattacharya}\ \emph {et~al.}(2017)\citenamefont {Bhattacharya}, \citenamefont {Metz},\ and\ \citenamefont {Zhou}}]{Bhattacharya:2017bvs}%
  \BibitemOpen
  \bibfield  {author} {\bibinfo {author} {\bibfnamefont {S.}~\bibnamefont {Bhattacharya}}, \bibinfo {author} {\bibfnamefont {A.}~\bibnamefont {Metz}}, \ and\ \bibinfo {author} {\bibfnamefont {J.}~\bibnamefont {Zhou}},\ }\href {\doibase 10.1016/j.physletb.2017.05.081} {\bibfield  {journal} {\bibinfo  {journal} {Phys. Lett. B}\ }\textbf {\bibinfo {volume} {771}},\ \bibinfo {pages} {396} (\bibinfo {year} {2017})},\ \bibinfo {note} {[Erratum: Phys.Lett.B 810, 135866 (2020)]},\ \Eprint {http://arxiv.org/abs/1702.04387} {arXiv:1702.04387 [hep-ph]} \BibitemShut {NoStop}%
\bibitem [{\citenamefont {Cocuzza}\ \emph {et~al.}(2024)\citenamefont {Cocuzza}, \citenamefont {Metz}, \citenamefont {Pitonyak}, \citenamefont {Prokudin}, \citenamefont {Sato},\ and\ \citenamefont {Seidl}}]{Cocuzza:2023oam}%
  \BibitemOpen
  \bibfield  {author} {\bibinfo {author} {\bibfnamefont {C.}~\bibnamefont {Cocuzza}}, \bibinfo {author} {\bibfnamefont {A.}~\bibnamefont {Metz}}, \bibinfo {author} {\bibfnamefont {D.}~\bibnamefont {Pitonyak}}, \bibinfo {author} {\bibfnamefont {A.}~\bibnamefont {Prokudin}}, \bibinfo {author} {\bibfnamefont {N.}~\bibnamefont {Sato}}, \ and\ \bibinfo {author} {\bibfnamefont {R.}~\bibnamefont {Seidl}} (\bibinfo {collaboration} {JAM}),\ }\href {\doibase 10.1103/PhysRevLett.132.091901} {\bibfield  {journal} {\bibinfo  {journal} {Phys. Rev. Lett.}\ }\textbf {\bibinfo {volume} {132}},\ \bibinfo {pages} {091901} (\bibinfo {year} {2024})},\ \Eprint {http://arxiv.org/abs/2306.12998} {arXiv:2306.12998 [hep-ph]} \BibitemShut {NoStop}%
\bibitem [{\citenamefont {Gupta}\ \emph {et~al.}(2018)\citenamefont {Gupta}, \citenamefont {Yoon}, \citenamefont {Bhattacharya}, \citenamefont {Cirigliano}, \citenamefont {Jang},\ and\ \citenamefont {Lin}}]{Gupta:2018lvp}%
  \BibitemOpen
  \bibfield  {author} {\bibinfo {author} {\bibfnamefont {R.}~\bibnamefont {Gupta}}, \bibinfo {author} {\bibfnamefont {B.}~\bibnamefont {Yoon}}, \bibinfo {author} {\bibfnamefont {T.}~\bibnamefont {Bhattacharya}}, \bibinfo {author} {\bibfnamefont {V.}~\bibnamefont {Cirigliano}}, \bibinfo {author} {\bibfnamefont {Y.-C.}\ \bibnamefont {Jang}}, \ and\ \bibinfo {author} {\bibfnamefont {H.-W.}\ \bibnamefont {Lin}},\ }\href {\doibase 10.1103/PhysRevD.98.091501} {\bibfield  {journal} {\bibinfo  {journal} {Phys. Rev. D}\ }\textbf {\bibinfo {volume} {98}},\ \bibinfo {pages} {091501} (\bibinfo {year} {2018})},\ \Eprint {http://arxiv.org/abs/1808.07597} {arXiv:1808.07597 [hep-lat]} \BibitemShut {NoStop}%
\end{thebibliography}%

\end{document}